\documentclass{aastex63}
\usepackage{amssymb,amsmath,graphicx,latexsym}
\usepackage{subfigure}
\shorttitle{Primordial black holes generated by fast-roll mechanism ...}
\shortauthors{Heydari and Karami}
\begin{document}
\title{Primordial black holes generated by fast-roll mechanism in non-canonical natural inflation}
\email{s.heydari@uok.ac.ir}
\email{kkarami@uok.ac.ir}
\author{Soma Heydari}
\affiliation{Department of Physics, University of Kurdistan, Pasdaran Street, P.O. Box 66177-15175, Sanandaj, Iran}
\author{Kayoomars Karami}
\affiliation{Department of Physics, University of Kurdistan, Pasdaran Street, P.O. Box 66177-15175, Sanandaj, Iran}

\begin{abstract}
In this work, a new fast-roll (FR) mechanism to generate primordial black holes (PBHs) and gravitational waves (GWs) in generalized non-canonical natural inflation is introduced. In this model, choosing a suitable function for non-canonical mass scale parameter $M(\phi)$ gives rise to produce a cliff-like region in the field evolution path. When inflaton rolls down the steep cliff, its kinetic energy during a FR stage increases in comparison with a slow-roll (SR) stage. Hence, seeds of PBH production are born in this transient FR stage. Depending on the position of the cliff, appropriate cases of PBHs for explaining total dark matter (DM), microlensing effects, LIGO-VIRGO events and NANOGrav 15 year data can be formed. The density spectrum of GWs related to one case of the model lies in the NANOGrav 15 year domain and behaves like $\Omega_{\rm GW_0}\sim f^{5-\gamma}$. The spectral index $\gamma=3.42$ for this case satisfies the NANOGrav 15 year constraint.
Moreover, regarding reheating considerations, it is demonstrated that PBHs are born in the radiation-dominated (RD) era. Furthermore, viability of the model in light of theoretical swampland criteria and observational constraints on cosmic microwave background (CMB) scales are illustrated.
\end{abstract}
\section{Introduction}
Recently, primordial black holes (PBHs) as  remarkable candidates for the cosmological budget of dark matter (DM) have attracted a lot of interests from scientific community \citep{Heydari:2024-a,Heydari:2024-b,Kamenshchik:2019,fu:2019,Dalianis:2019,
mahbub:2020,mishra:2020,Solbi-a:2021,Solbi-b:2021,Teimoori-b:2021,Laha:2019,Teimoori:2021,fu:2020,
Pi,Garcia-Bellido:2017,Heydari:2022,Heydari-b:2022,Rezazadeh:2021,Kawaguchi:2023,
Kawai:2021edk,Saridakis:2023,Clesse:2015,Kawasaki:2016,Braglia:2020,Motohashi:2017,Sayantan-4:2023,
Sayantan-6:2023,Ashrafzadeh:2023,Domenech:2021,
Domenech-2:2020,Drees:2021,Kawai:2022emp,Ahmed,
Solbi:2024,Cai:2023,Caravano:2024,Stamou:2021,Stamou:2024,Dimastrogiovanni:2024}.
PBHs with masses around $10M_\odot$ ($M_\odot$ expresses the solar mass) also could be considered as possible sources of  gravitational waves (GWs) ferreted out by LIGO-Virgo teamwork \citep{Abbott:2016-a,Abbott:2016-b,Abbott:2017-a,Abbott:2017-b,Abbott:2017-c}. Into the bargain, lighter PBHs with masses around the Earth mass could be used to explain  microlensing events in OGLE data \citep{OGLE-1,OGLE-2}. Lately, pulsar timing array EPTA, PPTA and CPTA as well as NANOGrav teamwork have succeed to trace low-frequency GWs  \citep{NANOGrav-1,NANOGrav-2,NANOGrav-3,NANOGrav-4,NANOGrav-5,
EPTA-1,EPTA-2,EPTA-3,EPTA-4,EPTA-5}. Interestingly enough, the emitted GWs from PBHs formation could be contemplated as promising origins of the NANOGrav signals
\citep{Franciolini:2023,Vaskonen,Luca:2021,Kohri:2021,Sugiyama:2021,Inomata:2021,Atal:2021,Kawasaki:2021}. Moreover, PBHs are likely candidate for formation of super-massive black holes in the center of galaxies \citep{Hasegawa:2021,Kawasaki:2019,Kitajima:2020,Shinohara:2021}.
Generally, PBHs are born from sizable density fluctuations, when they return to the horizon in the radiation-dominated (RD) era. These fluctuations originate from scalar fluctuations produced in the inflation epoch. Basically, in order to produce seeds of PBHs in the inflationary epoch, it is required to have a proper mechanism for enhancing the power spectrum of scalar perturbations to ${\cal P}_{\cal R}\sim{\cal O}(10^{-2})$ on small scales with respect to that on cosmic microwave background (CMB) scale.

Mostly, PBHs models include an inflationary epoch composed of three sequential stages so called, first slow-roll (SR) stage, intermediate ultra-slow-roll (USR) stage and final SR stage. Usually, in the first SR stage, the CMB anisotropies are produced and leave the horizon. In this stage, the first and second SR parameters ($\varepsilon_1$ and  $\varepsilon_2$) are very smaller than one to guarantee the SR conditions $\{\varepsilon_1,\varepsilon_2\}\ll1$. The inflaton in the USR stage, owing to high friction,  decelerates and its kinetic energy decreases in comparison with the SR stage. Also,  severe reduction in  the $\varepsilon_1$ value takes place and simultaneously, the $\varepsilon_2$ parameter after reaching a large negative value takes large positive values higher than one. So, the SR condition is violated by $\varepsilon_2$ in the USR stage and subsequent to that, the amplitude of  ${\cal P}_{\cal R}$ boosts to ${\cal O}(10^{-2})$, which is essential to produce PBHs \citep{Heydari:2024-a,Heydari:2024-b,fu:2019,Dalianis:2019,
mahbub:2020,mishra:2020,Solbi-a:2021,Solbi-b:2021,Teimoori-b:2021,Laha:2019,Teimoori:2021,fu:2020,
Heydari:2022,Heydari-b:2022,Solbi:2024,Ashrafzadeh:2023}. After the USR stage, the inflaton enters the final SR stage which realizes the end of inflation. In this final stage, the values of $\varepsilon_1$, $\varepsilon_2$ and kinetic energy of the inflaton almost return to their values from the first SR stage.

But the USR mechanism is not the only case for PBHs generation. In recent years, some inflationary models have been introduced to explain PBHs generation, which containing special form of potentials in the presence of downward step like cliff on small scale far from the CMB scale \citep{Inomata:2021-b,Ozsoy:2021}. In these works, it is demonstrated that, amplification of the scalar power spectrum to around 7 order of magnitude through a transient fast-roll (FR) mechanism during the stage of cliff passing is possible. In these models, inflationary era is composed of the first SR stage to produce CMB anistropies, the intermediate FR stage to produce PBHs and the final SR to end of the inflation. In the FR stage, the inflaton experiences increase in its speed, kinetic energy and value of $\varepsilon_1$ parameter. Same as the USR stage, the SR condition in the FR stage is violated via the $\varepsilon_2$ parameter. In fact, particle production as a result of non-adiabatic transition between SR and FR stage in the cliff position, gives rise to perturbation magnification and born PBHs seeds. Hitherto, FR mechanism is established in canonical standard model of inflation through altering the potential to realize a cliff-like region in the path of field evolution \citep{Inomata:2021-b,Ozsoy:2021}.

It is known that, one of the successful generalization of the standard model of inflation to revive the steep potentials and produce PBHs and GWs is power-law non-canonical (PLNC) model containing Lagrangian ${\cal L}(X,\phi)=X^{\alpha}-V(\phi)$ ($X$ expresses canonical kinetic energy)  \citep{Unnikrishnan:2012,Rezazadeh:2015,Saridakis:2023,Heydari:2024-a,Heydari:2024-b,Mishra:2018,Mishra:2021,Unnikrishnan:2013,safaii:2024}. It is obvious that, diversion of the canonical standard inflationary model is defined by  $\alpha>1$ ($\alpha=1$ corresponds to canonical Lagrangian).
In \citep{Heydari:2024-a}, PBHs and GWs formation from the quartic potential in generalized power-law non-canonical (GPLNC) inflationary model have been investigated. In the GPLNC setup, non-canonical mass scale parameter $M(\phi)$ is chosen as a general function of the scalar field rather than being constant as in the PLNC model.

The aim of this study is probing generation of PBHs and  GWs from natural potential in the GPLNC setup through the FR mechanism. It is tried to utilize the dependency of $M$ to the scalar field to produce a FR span on a cliff in the field evolution pass to generate PBHs. Using the reheating consideration, we attempt to show that PBHs are born in the RD era.
To this end, in Sec. \ref{sec2}, a brief review of GPLNC model is presented. In Sec. \ref{sec3}, feasibility of the model theoretically and observationally is checked and used method to produce PBHs is explained. In Sec. \ref{sec4}, reheating computation is described. Thence, the spectra of PBHs mass in Sec. \ref{sec5} and secondary GWs in Sec. \ref{sec6} are elucidated. Lastly, Sec. \ref{sec7} is earmarked as conclusions of the work.
\section{Review of the GPLNC model }\label{sec2}
\subsection{Background evolution}\label{bg}
This section is started with the ensuing action
\begin{equation}\label{action}
S=\int{\rm d}^{4}x \sqrt{-g} \left[M_{P}^2\frac{R}{2}+{\cal L}(X,\phi)\right] ,
\end{equation}
therein, $g$, $M_{P}=1/\sqrt{8\pi G}$, $R$ and $\mathcal{L}(X,\phi)$ are respectively, determinant of the metric tensor $g_{\mu\nu}$, the reduced Planck mass, the Ricci scalar and the Lagrangian density. Regarding the generality of the Lagrangian density function to the scalar field $\phi$ and  kinetic energy $X\equiv\frac{1}{2}g_{\mu\nu}\partial^\mu \phi \partial^\nu \phi$ \citep{Unnikrishnan:2012,Rezazadeh:2015,Saridakis:2023,Heydari:2024-a,Heydari:2024-b,Mishra:2018,
Mishra:2021,Unnikrishnan:2013,safaii:2024}, we are interested in the GPLNC form of that as  \citep{Heydari:2024-a}
\begin{equation}\label{Lagrangian}
{\cal L}(X,\phi) = X\left(\frac{X}{M^{4}(\phi)}\right)^{\alpha-1} - V(\phi),
\end{equation}
wherein, $\alpha$ with no dimension defines diversion from canonicity. It means that, if $\alpha=1$, the Lagrangian (\ref{Lagrangian}) regains its canonical form. The non-canonical mass scale parameter $M(\phi)$ with mass dimension could be specified as a generic function of the scalar field $\phi$. It is notable that, the conventional form of the Lagrangian of the PLNC model with constant $M$ could be recovered through a suitable field redefinition for the Lagrangian (\ref{Lagrangian}) \citep{Heydari:2024-a}. Furthermore, the inflationary potential is denoted by $V(\phi)$ in the Lagrangian (\ref{Lagrangian}).
In the following, utilizing the Lagrangian density (\ref{Lagrangian}), one can compute the energy density $\rho_{\phi}$ and pressure $p_{\phi}$ for the scalar field as
\begin{eqnarray}
{\rho _\phi } &=&2X {\cal L}_{,X} - {\cal
L}= \left( {2\alpha - 1}
\right)X{\left( {\frac{X}{{{M^4(\phi)}}}} \right)^{\alpha  - 1}} +
V(\phi), \label{eq:rho}
\\
{p_\phi } &=& {\cal L}=X{\left( {\frac{X}{{{M^4(\phi)}}}} \right)^{\alpha  - 1}} - V(\phi ),\label{eq:p}
\end{eqnarray}
where, $(_{,X})$ means partial derivative with respect to $X$. Homogeneity and isotropy of the Universe are expressed via the flat Friedmann-Robertson-Walker (FRW) metric as $g_{{\mu}{\nu}}={\rm diag}\Big(1, -a^{2}(t), -a^{2}(t), -a^{2}(t)\Big)$,
wherein $a(t)$ indicates scale factor as a function of cosmic time $t$.
Taking variation of the action (\ref{action}) against the metric $g_{{\mu}{\nu}}$  results in the Friedmann equations as
\begin{eqnarray}
\label{eq:Friedmann}
H^{2} &=&\frac{\rho _\phi}{3 M_{P}^2}= \frac{1}{3 M_{P}^2}\left[\left(2\alpha-1\right)X\left(\frac{X}{M^{4}(\phi)}\right)^{\alpha-1} +\;
V(\phi)\right],\label{eq: FR-eqn1}\\
\dot{H} &=&\frac{-(\rho _\phi+p_{\phi})}{2 M_{P}^2} =-\frac{\alpha}{M_{P}^2} X\left(\frac{X}{M^4(\phi)}\right)^{\alpha -1},
\label{eq: FR-eqn2}
\end{eqnarray}
therein $H\equiv \dot{a}/a $ signifies the Hubble parameter (dot means derivative versus $t$). Likewise, taking variation of the action (\ref{action}) against the scalar field $\phi$  results in the field evolution equation as
\begin{equation}
\label{eq:KG}
\ddot \phi  + \frac{{3H\dot \phi }}{{2\alpha  - 1}}-2\left(1-\frac{1}{\alpha}\right)\left(\frac{M_{,\phi}}{M(\phi)}\right)\dot\phi^2
+\frac{1}{\alpha} \left( {\frac{{V_{,\phi }}}{{2\alpha  - 1}}}
\right){\left( {\frac{{2{M^4(\phi)}}}{{{{\dot \phi }^2}}}} \right)^{\alpha
- 1}} = 0,
\end{equation}
in which $(_{,\phi})$ means derivative versus $\phi$. Thenceforth, the 1st and 2nd Hubble SR parameters are given by
\begin{equation}\label{SRP}
  \varepsilon_{1} \equiv -\frac{\dot H}{H^2}, \hspace{.5cm}  \varepsilon_{2} \equiv \frac{\dot{\varepsilon_{1}}}{ H\, \varepsilon_{1}}.
\end{equation}
In the SR domain $\{ \varepsilon_{1},\varepsilon_{2}\}\ll 1$, the potential energy prevails over the kinetic energy.
It is notable that, conflation of Eqs. (\ref{eq:Friedmann})-(\ref{eq:KG}) and utilizing the 1st SR parameter beside the $e$-folds number definition $dN=Hdt$, the equation of field evolution (\ref{eq:KG}) is rewritten as ensuing form
\begin{equation}
\label{eq:KG-NC}
\phi_{,NN}+\left[\frac{V_{,\phi}}{2V\varepsilon_{1}}\left(\frac{3}{2\alpha-1}-
\frac{\varepsilon_{1}}{\alpha}\right)-\frac{2M_{,\phi}}{M(\phi)}\left(1-\frac{1}{\alpha}\right)
\right]\phi_{,N}^{2}+\left[\frac{3}{2\alpha -1} - \varepsilon_{1} \right]\phi_{,N}=0,
\end{equation}
therein $(_{,N})$ and $(_{,NN})$ designate the 1st and 2nd derivative versus $e$-folds number $N$.
\subsection{Evolutions of the scalar and tensor perturbations}
Subsequent to analyses of \citep{Garriga:1999,Unnikrishnan:2012} in NC setup, the evolutions of the scalar and tensor perturbations are governed by the Mukhanonv-Sasaki (MS) equations, which are derived from expanding the action (\ref{action}) to the second order in perturbations. In this framework, the MS equation for the curvature perturbations  ${\cal R}$ is given by
\begin{equation}\label{eq:MS}
 u^{\prime\prime}_{k}+\left(c_{s}^2 k^2-\frac{z^{\prime\prime}}{z}\right)u_k=0,
\end{equation}
where, $(~^\prime)$ means derivative versus conformal time $\eta\equiv\int {dt/a(t)}$ and
\begin{equation}\label{eq:z}
 u_k\equiv z {\cal R}_k, \hspace{.9cm}z \equiv \frac{a\,\sqrt{\rho_{_{\phi}}+ p_{_{\phi}}}}{c_{_s}H},\hspace{.9cm}
 c_s^2\equiv\frac{{\cal L}_{,X}}{2X{\cal L}_{,XX}+{\cal L}_{,X}}.
\end{equation}
Here, $c_s$ is the sound speed of the scalar perturbations. Thus, the power spectrum of the curvature perturbations is acquired from the solution of the MS equation (\ref{eq:MS}) as
\begin{equation}\label{eq:PsMS}
{\cal P}_{\cal R}\equiv\frac{k^3}{2\pi^2}\left|{\frac{u^2_k}{z^2}}\right|.
\end{equation}
In the SR domain when the comoving wavenumber $k$ crosses the sound horizon $(c_{s}k=aH)$, the ${\cal P}_{\cal R}$ is approximated as \citep{Garriga:1999}
\begin{equation}\label{eq:Ps-SR}
{\cal P}_{\cal R}=\frac{H^2}{8 \pi ^{2}M_{P}^{2} c_{s} \varepsilon_{1}}\Big|_{c_{s}k=aH}.
\end{equation}
The observational restriction on the curvature power spectrum
at pivot scale $(k_{*}=0.05~\rm Mpc^{-1})$ enforced by Planck collaboration is
 $ {\cal P}_{\cal R}(k_{*})\simeq 2.1 \times 10^{-9}$ \citep{akrami:2018}.
Regarding the Lagrangian (\ref{Lagrangian}), the $c_s$ in our setup is obtained as
\begin{equation}
\label{eq:cs}
c_s =(2\alpha  - 1)^{-1/2}.
\end{equation}
So as to preclude the model from classical instability, it is essential to have $0<c_s^2\le 1$ which leads to a lower bound on the $\alpha$ parameter as $\alpha\ge 1$. Moreover, using the curvature power spectrum (\ref{eq:Ps-SR}), the scalar spectral index $n_{s}$ can be derived with respect to the SR parameters as
\begin{align}\label{eq:ns-SR}
n_s-1\equiv \frac{d\ln{\cal P}_{\cal R}}{d\ln k}\simeq -2\varepsilon_1-\varepsilon_2.
\end{align}
The observational bound on the scalar spectral index through the recent data of Planck 2018 is
$n_s= 0.9653 \pm 0.0041$ (TT, TE, EE+lowE+lensing+BK18+BAO, 68\%  CL)  \citep{akrami:2018,Paoletti:2022}.

Likewise, the MS equation governing the tensor perturbations in this setup is given by
\begin{equation}\label{eq:tMS}
 \upsilon^{\prime\prime}_{k}+\left( k^2-\frac{a^{\prime\prime}}{a}\right)\upsilon_k=0,
\end{equation}
wherein, $\upsilon_k\equiv a {h}_k$.
So, the power spectrum of tensor perturbations is calculated from the solution of the tensor MS equation (\ref{eq:tMS}) as follows
\begin{equation}\label{eq:PtMS}
{\cal P}_{t}\equiv\frac{k^3}{\pi^2}\left|
{\frac{{\upsilon_k}^2}{a^2}}\right|,
\end{equation}
which, in the SR domain at  $k=aH$, is approximated as \citep{Garriga:1999}
\begin{equation}\label{eq:Pt-SR}
{\cal P}_{t}=\frac{2H^2}{\pi ^{2}M_{P}^2}\Big|_{k=aH}.
\end{equation}
The tensor-to-scalar ratio $r$ in terms of the 1st SR parameter is attained from   (\ref{eq:Ps-SR}) and  (\ref{eq:Pt-SR}) as
\begin{equation}
\label{eq:r}
r\equiv\frac{{\cal P}_t}{{\cal P}_{\cal R}}\simeq16 c_s \varepsilon_1.
\end{equation}
The recent upper limit on the $r$, established by Planck and BICEP/Keck 2018 data, is $r<0.036$ (TT, TE, EE +LowE +Lensing +BK18+BAO,  95\% CL) \citep{BK18:2021}.
\section{Fast-Roll mechanism and feasibility of the model }\label{sec3}
The aim of this section is showing that, using the FR mechanism in the GPLNC model, the required enhancement in the curvature power spectrum to produce PBHs seeds can occur. At the same time, this enhancement on small scales should not ruin the feasibility of the model on CMB scale in light of theoretical and observational constraints.
Therefore, the non-canonical mass scale parameter $M(\phi)$ of the Lagrangian (\ref{Lagrangian}) is defined as a twofold function of $\phi$ as follows
\begin{equation}\label{eq:M}
M(\phi) = M_0\Big[1+\epsilon(\phi)\Big],
\end{equation}
where
\begin{equation}\label{eq:bump}
\epsilon(\phi) =\omega\cosh^{-2}\Big(\frac{\phi-\phi_{c}}{b}\Big).
\end{equation}
In the 1st part of Eq. (\ref{eq:M}), the parameter $M_0$ with mass dimension, affects the consistency of the model with the CMB data on the large scales. In the second part, the dimensionless function $\epsilon(\phi)$ is employed to produce a local steep cliff in the field evolution path. For this purpose, a piked function like Eq. (\ref{eq:bump}) is selected for $\epsilon(\phi)$, which illustrates a local peak at $\phi=\phi_{c}$ with width $b$ and elevation $\omega$. Here, $\phi_c$ and $b$ have mass dimensions whereas $\omega$ is dimensionless. From Eq. (\ref{eq:bump}), it is obvious that the piked function  $\epsilon(\phi)$ for $\phi\neq\phi_{c}$ gradually dies out, i.e., $\epsilon(\phi)\ll 1$. Thus, $M(\phi)$ in the GPLNC setup reverts to the constant $M_0$ in the conventional PLNC setup.
The function $\epsilon(\phi)$ in the GPLNC framework, has the ability to produce a local steep cliff with the altitude of $\omega$, in $\phi=\phi_{c}$, on the field evolution path. When the inflaton reaches the cliff position, rolls down faster and its kinetic energy enhances in comparison with the SR domain (see plots of Fig. \ref{fig:bg}). Hence, a transient FR stage is produced in the proximity of $\phi=\phi_{c}$ and subsequent to that the curvature power spectrum enhances to required value for generating PBHs seeds. In fact, a non-adiabatic transition from the SR stage to the FR stage gives rise to particle production and enhance the scalar perturbation to produce PBHs. Therefore, it is necessary to regulate three parameters $\{\omega,\phi_c,b\}$ of the $\epsilon(\phi)$ function to have a suitable FR stage to generate PBHs seeds on the small scales.

In the following, natural inflationary potential is chosen for the model
\begin{equation}\label{eq:qV}
V(\phi)=\Lambda^{4}\Bigg[1+\cos\Bigg(\frac{\phi}{f}\Bigg)\Bigg],
\end{equation}
where, both $\Lambda$ and $f$ have dimension of mass. The prognostications of the natural potential on the CMB scale are not consistent with the latest data of Planck 2018 in the standard model of inflation \citep{akrami:2018,BK18:2021}.  Accordingly, we strive to rectify its shortcomings on the CMB scale using the non-canonical parameters $\{\alpha,M_0\}$ of the GPLNC framework. The $\Lambda$ and $M_0$ parameter are associated together through the SR relation of the scalar power spectrum (\ref{eq:Ps-SR}). Hence, fixing $\Lambda=0.01 M_P$ to guarantee the correct energy scale of inflation, the parameter $M_0$ is attained from the CMB normalization confinement ${\cal P}_{\cal R}(k_{*})\sim2.1 \times 10^{-9}$ at pivot scale $(k_{*}=0.05~\rm Mpc^{-1})$ \citep{akrami:2018}. The $\alpha$ parameter is responsible to cure the prognostications of the natural potential for $r$ and $n_s$ in the GPLNC setup. Table \ref{tab1} is  allotted to common fixed  parameters for all cases of the model, while Table \ref{tab2} contains the regulated parameter sets for each case. The computed values for $n_s$,  $r$ and related parameters to PBHs are listed in Table \ref{tab3}.

Generally, on the small scales, among the model parameters, $\omega$ and $\phi_c$ need more fine-tuning to gain meaningful results in PBHs generation. The parameter $\omega$ specifies the altitude of the peak of the scalar power spectrum and consequently abundance of the resulted PBHs. The parameter $\phi_c$ defines the peak position in the scalar power spectrum and subsequently the location of PBHs mass spectrum, so it should be fine tuned to produce PBHs with favorable mass. The parameter $b$ specifies the width of the peak of the scalar power spectrum and the mass spectrum of the generated PBHs. On the large (CMB) scale, we increase the non-canonical parameter $\alpha$ to decrease the tensor-to-scalar ratio $r$ and duration of the inflationary era is affected by tuning the parameter $M_0$.

\begin{table*}[ht!]
  \centering
  \caption{Common fixed parameters for cases A, B, C, D and F.}
\begin{tabular}{ccccc}
  \hline
 $\#$&\qquad $\alpha$\qquad &\qquad $b/M_{P}$\qquad&\qquad $f/M_{P}$\qquad&\qquad $\Lambda/M_{P}$\qquad\\[0.5ex] \hline\hline
  All cases& \qquad$20$\qquad  &\qquad$1.2\times10^{-2}$ \qquad&\qquad$2$\qquad& \qquad $0.01$\qquad\\[0.5ex] \hline
\end{tabular}
 \label{tab1}
\end{table*}
\begin{table*}[ht!]
  \centering
  \caption{Tuned parameters for all cases of the model. }
\begin{tabular}{cccccc}
  \hline
 $\#$ &\qquad $\omega$\qquad & \qquad$\phi_{c}/M_{P}$\qquad&\qquad$M_0/M_{P}$\qquad\\[0.5ex] \hline\hline
  Case A& \qquad$14.895$\qquad &\qquad$5.809$ \qquad&\qquad$2.33\times10^{-4}$\qquad\\[0.5ex] \hline
  Case B&\qquad $15.220$\qquad &\qquad$5.740$\qquad &\qquad$2.33\times10^{-4}$\qquad\\ \hline
  Case C&\qquad$15.440$\qquad & \qquad$5.692$\qquad&\qquad$2.32\times10^{-4}$\qquad\\ \hline
  Case D&\qquad$13.100$\qquad & \qquad$5.709$\qquad&\qquad$2.32\times10^{-4}$\qquad\\ \hline
  Case F&\qquad$14.385$\qquad & \qquad$5.860$\qquad&\qquad$2.46\times10^{-4}$\qquad\\ \hline
\end{tabular}
 \label{tab2}
\end{table*}
\begin{table*}[ht!]
  \centering
  \caption{Computed quantities corresponding to the cases of Tables \ref{tab1}-\ref{tab2} for $n_{s}$,  $r$, ${\cal P}_{ \cal R}^\text{peak}$, $k_{\text{peak}} $,   $f_{\text{PBH}}^{\text{peak}}$ and $M_{\text{PBH}}^{\text{peak}}$. The values of $n_{s}$ and $r$ are evaluated at horizon intersecting $e$-folds number $N_{*}=0$ by the pivot scale.}
\begin{tabular}{ccccccc}
  \hline
   $\#$ & \quad $n_{s}$\quad &\quad $r$\quad &\quad$ {\cal P}_{\cal R}^{\text{peak}}$\quad &\quad$k_{\text{peak}}/\text{Mpc}^{-1}$\quad& \quad$f_{\text{PBH}}^{\text{peak}}$\quad& \quad$M_{\text{PBH}}^{\text{peak}}/M_{\odot}$\quad\\ \hline\hline
  Case A &\quad0.9686\quad  &\quad0.020\quad& \quad0.036\quad & \quad$3.06\times10^{12}$ \quad&\quad 0.9963\quad &\quad$2.52\times10^{-13}$\quad\\ \hline
 Case B &\quad 0.9686\quad  & \quad 0.020\quad &\quad0.043\quad&\quad  $3.59\times10^{8}$ \quad&\quad0.0182\quad &\quad $1.83\times10^{-5}$ \quad\\ \hline
 Case C &\quad0.9684\quad  & \quad0.020\quad & \quad0.053\quad&\quad$3.63\times10^{5}$\quad &\quad 0.0016\quad &\quad$17.88$ \quad\\ \hline
 Case D &\quad0.9684\quad  & \quad0.020\quad & \quad0.013\quad & \quad$6.90\times10^{6}$\quad &\quad$\sim 0$\quad &\quad$0.0496$ \quad\\ \hline
Case F &\quad0.9644\quad  & \quad0.022\quad & \quad0.034\quad & \quad$1.36\times10^{13}$\quad &\quad$0.9853$\quad &\quad$1.27\times10^{-14}$ \quad\\ \hline
\end{tabular}
\label{tab3}
\end{table*}
\subsection{Background evolutions}
\begin{figure*}
\begin{minipage}[b]{1\textwidth}
\centering
\subfigure[\label{phi}]{ \includegraphics[width=.472\textwidth]%
{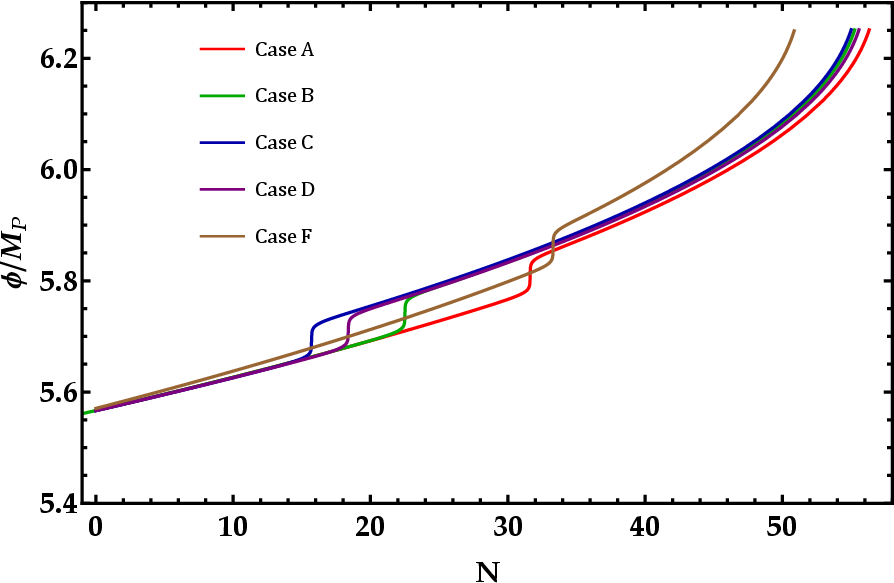}}\hspace{.1cm}
\subfigure[\label{h}]{ \includegraphics[width=.499\textwidth]%
{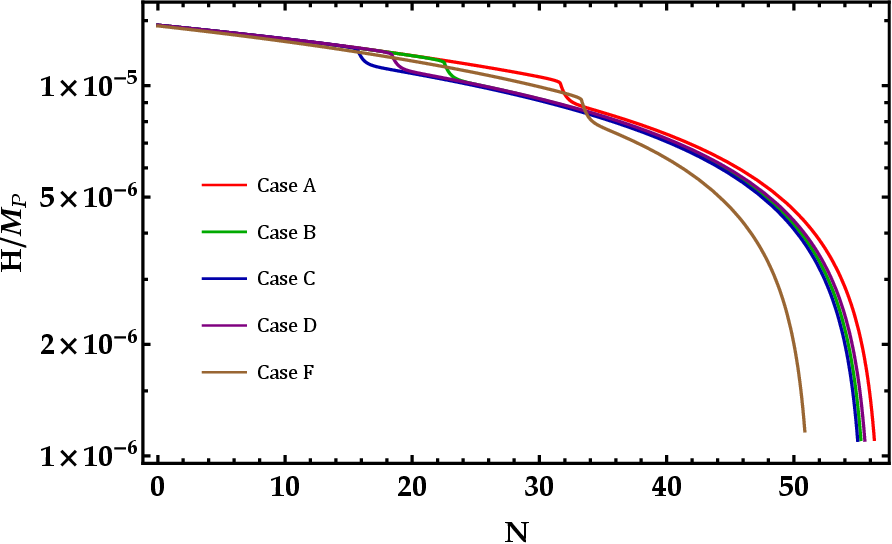}}\hspace{.1cm}
\subfigure[\label{dphi}]{ \includegraphics[width=.485\textwidth]%
{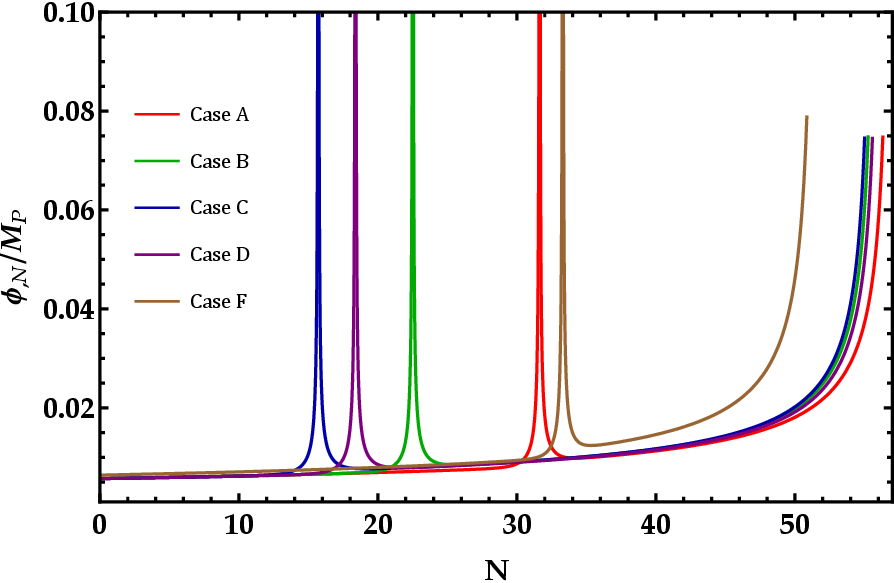}}\hspace{.1cm}
\subfigure[\label{e1}]{\includegraphics[width=.485\textwidth]%
{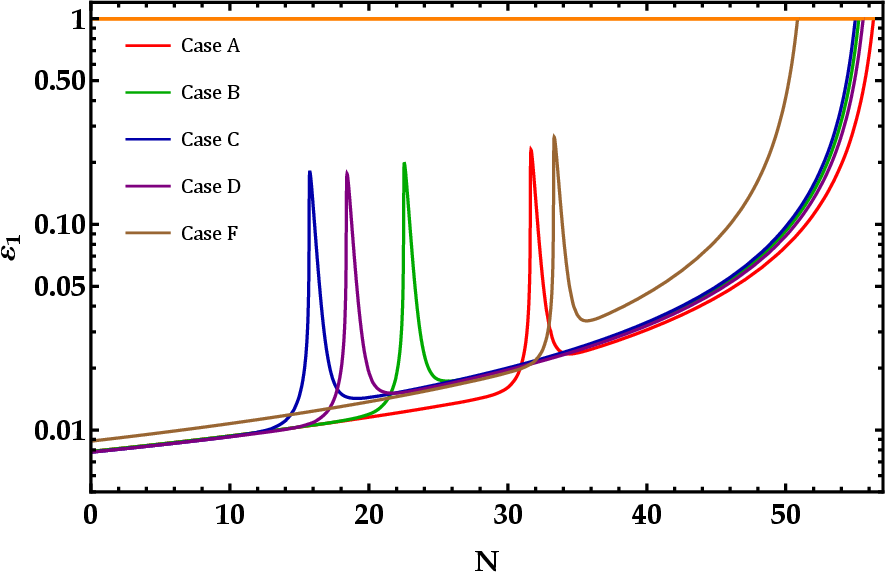}} \\\hspace{.1cm}
\centering
\subfigure[\label{e2}]{ \includegraphics[width=.5\textwidth]%
{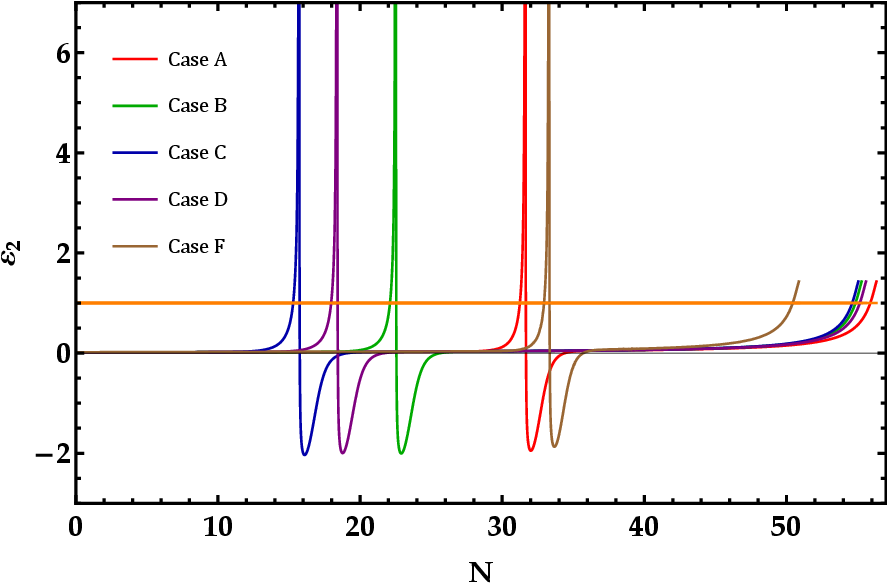}}\hspace{.1cm}
\end{minipage}
\caption{Variations of (a) the scalar field, (b) the Hubble parameter, (c) the field derivative, (d) the first SR parameter and (e) the second SR parameter against the $e$-folds number $N$ for cases A (red curves), B (green curves), C (blue curves), D (purple curves) and F (brown curves).}
\label{fig:bg}
\end{figure*}

So as to compute the results of Table \ref{tab3}, we need to know the dynamics of background.
Therefore, using the numerical method, the equation of field evolution (\ref{eq:KG-NC}) and the 2nd Friedmann equation (\ref{eq: FR-eqn2}) are solved at once. Thence, the variation of scalar field $\phi$, Hubble parameter $H$, field derivative $\phi_{,N}$, the 1st and 2nd slow-roll parameters ($\varepsilon_1$ and $\varepsilon_2$) versus the $e$-folds number $N$ for all cases of the model are computed and represented in Fig. \ref{fig:bg}.

It is inferable from Fig. \ref{fig:bg} that (i) the inflationary era persists for $\Delta N$ $e$-folds number ($\Delta N=N_{\rm end}-N_{*}$), between the horizon crossing moment by the pivot scale ($N_{*}=0$) and the end of inflation ($N_{\text{end}}$).  (ii) The end of inflation occurs through $\varepsilon_{1}=1$ for each case (Fig. \ref{e1}). (iii) The FR stage during a steep cliff in the field evolution pass in $\phi=\phi_c$ position is generated for each case of the model (Fig. \ref{phi}). (iv) In the FR stage, the speed of the inflaton increases (Fig. \ref{dphi}). (v) The $\varepsilon_{1}$ parameter experiences a transient peak during the FR stage (see Fig. \ref{e1}). (vi) At the same time, the $\varepsilon_2$ parameter in the FR stage increases to large positive values and crosses $\varepsilon_2=1$, momentarily (see Fig. \ref{e2}). Hence the SR condition in the FR stage is broken by $\varepsilon_2$.
(vii) Although the SR conditions $\{ \varepsilon_{1},\varepsilon_{2}\}\ll 1$ are broken in the FR stage, they are established around the CMB scale $N_{*}=0$ (Figs. \ref{e1} and  \ref{e2}).

Thus, computing the $n_s$ and $r$ via Eqs. (\ref{eq:ns-SR}) and (\ref{eq:r}), governed by the SR approximation, are permitted. According to the present results in Table \ref{tab3}, the values of $n_{s}$ for all cases of the model lie inside the 68\% CL of the permitted data of Planck 2018 (TT, TE, EE+lowE+lensing+BK18+BAO) \citep{akrami:2018}. Moreover, $r\sim0.02$ for each case of the model is consistent with the imposed upper bound on $r$ by Planck and BICEP/Keck 2018 data ($r<0.036$ at 95$\%$ CL) \citep{BK18:2021}.
As a consequence, the GPLNC framework have succeeded to amend the observational predictions of the natural potential.

\begin{figure*}
\begin{minipage}[b]{1\textwidth}
\subfigure[\label{swphi}]{ \includegraphics[width=.48\textwidth]%
{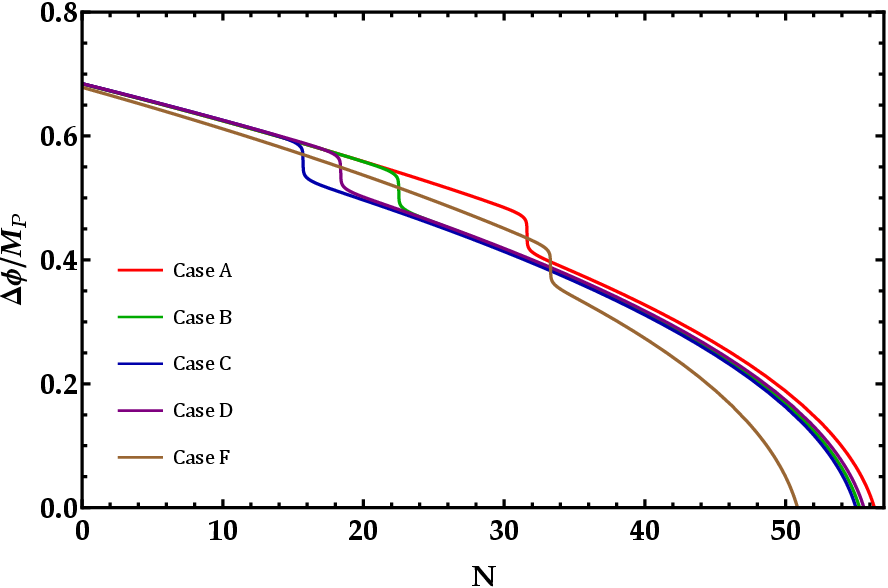}}\hspace{.1cm}
\subfigure[\label{swdv}]{ \includegraphics[width=.48\textwidth]%
{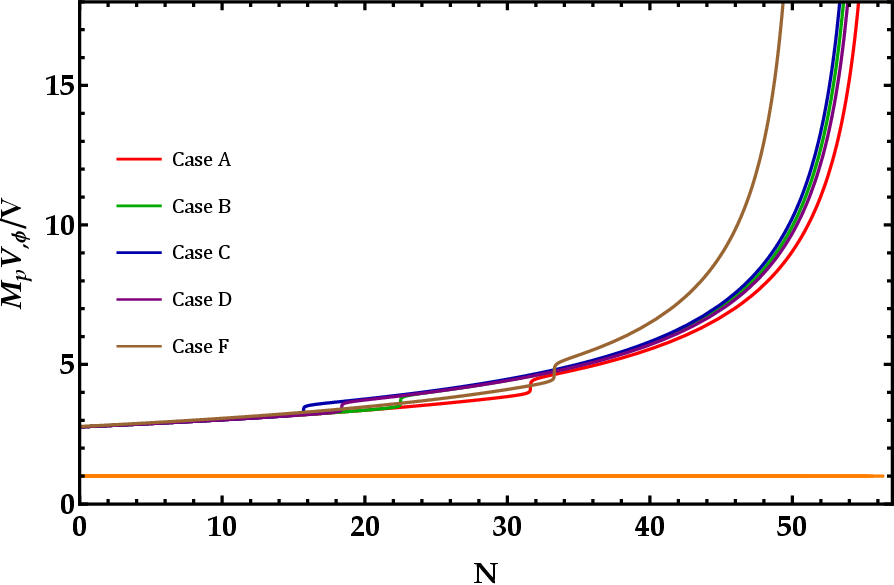}}\hspace{.1cm}
\end{minipage}
\caption{Establishment of the swampland criteria for cases of the model encompass (a) the distance conjecture $\Delta\phi/M_P<c_1$ and (b) the de-Sitter conjecture $M_P|V_{,\phi}/V|>c_2$.}
\label{fig:s}
\end{figure*}

In the following, we are interested in checking the swampland criteria for the model.
The swampland criteria encompass two conjectures to wit  distance conjecture and de-Sitter conjecture  \citep{Garg:2019,Ooguri:2019,Kehagias:2018}, which arise out of the string theory.
The distance conjecture imposes an upper limit on the field variation like $\Delta\phi/M_P<c_1$, while the de Sitter conjecture imposes a lower limit on the potential gradient like $M_P|V_{,\phi}/V|>c_2$. In these conjectures $c_1$ and $c_2$ are constants of order one. For the first conjecture, the field variation $\Delta\phi=\phi(N)-\phi(N_{\rm end})$ is evaluated and displayed in Fig. \ref{swphi} for each case of the model. Apropos to the second conjecture, Fig. \ref{swdv} illustrates the gradient of the inflationary potential $|V_{,\phi}/V|$ for cases of the model. The graphs of Fig. \ref{fig:s} indicate the validity of our model in light of the swampland criteria.

\subsection{Evolutions of the scalar and tensor perturbations}
\begin{figure*}
\begin{minipage}[b]{1\textwidth}
\subfigure[\label{ps}]{ \includegraphics[width=.495\textwidth]%
{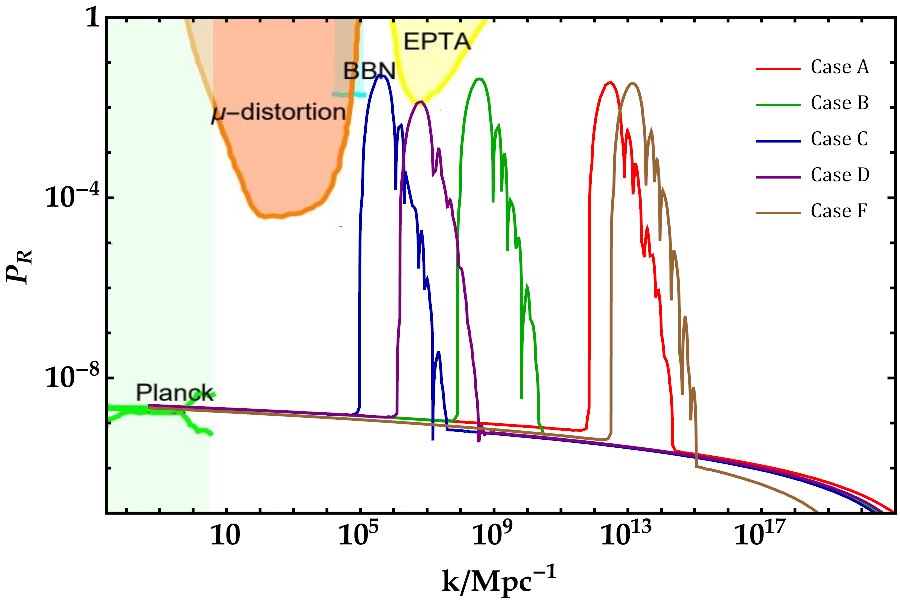}}\hspace{.1cm}
\subfigure[\label{pt}]{ \includegraphics[width=.485\textwidth]%
{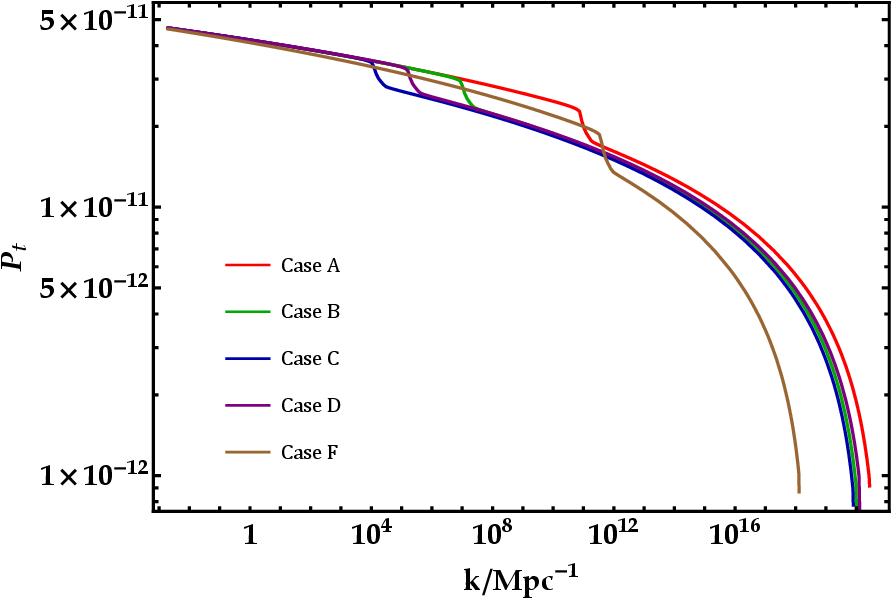}}\hspace{.1cm}
\end{minipage}
\caption{(a) The curvature and (b)  tensor power spectra with regard to the comoving wavenumber $k$ for Cases A (red curves), B (green curves), C (blue curves), D (purple curves) and F (brown curves). The colored zones are prohibited through the CMB observations \citep{akrami:2018} (light-green),  PTA observations \citep{Inomata:2019-a} (yellow), the impact on the ration of neutron-to-proton in the course of Big Bang Nucleosynthesis (BBN) \citep{Inomata:2016,Nakama:2014,Jeong:2014} (cyan), and the $\mu$-distortion of CMB \citep{Fixsen:1996,Chluba:2012} (orange).}
\label{fig:ps}
\end{figure*}

In what follows, after analyzing the background evolution, we need to know the evolutions of the scalar and tensor power spectra. At first, we solve the MS equation (\ref{eq:MS}) in numerical way to find the precise value of the curvature power spectrum from Eq. (\ref{eq:PsMS}). Thence, we plot the outcomes in Fig. \ref{ps} with regard to wavenumber $k$ for all cases of Table \ref{tab2} in addition to the observational restrictions. Also the numerical values of the peaks of the scalar power spectra (at $\phi=\phi_c$) and their related wavenumbers are listed in Table \ref{tab3} for all cases. Fig. \ref{ps} indicates that (i) the scalar power spectra adjacent to CMB scale $k_*\sim0.05~ \rm Mpc^{-1}$ for all cases satisfy the Planck confinement $ {\cal P}_{\cal R}(k_{*})\simeq 2.1 \times 10^{-9}$ \citep{akrami:2018}. (ii) The scalar power spectra respecting all cases experience large peaks with heights of order $10^{-2}$ adjacent to cliff positions ($\phi=\phi_c$) in the FR stages which is necessary to produce PBH seeds.

So as to compute the tensor power spectra for cases of our model, we solve the tensor MS equation (\ref{eq:tMS}) in numerical way. Thence, the precise values of the tensor power spectra are evaluated by Eq. (\ref{eq:PtMS}). The numerical outcomes for ${\cal P}_t$ are schemed with regard to wavenumber in Fig. \ref{pt}. This figure indicates that (i) on the CMB scale $k_*\sim0.05~ \rm Mpc^{-1}$, the tensor-to-scalar ratio $r={\cal P}_t(k_{*})/{\cal P}_{\cal R}(k_{*})$ results in the consistent values with the computed results of Table \ref{tab3} for $r$ through Eq. (\ref{eq:r}). (ii) The cliff-like region in the plot of the tensor power spectrum for each case originates from the cliff-like region in the Hubble parameter during the FR stage (see Fig. \ref{h}).
\section{Reheating consideration}\label{sec4}
\begin{figure*}
\begin{minipage}[b]{1\textwidth}
\vspace{-1.8cm}
\subfigure{ \includegraphics[width=.490\textwidth]%
{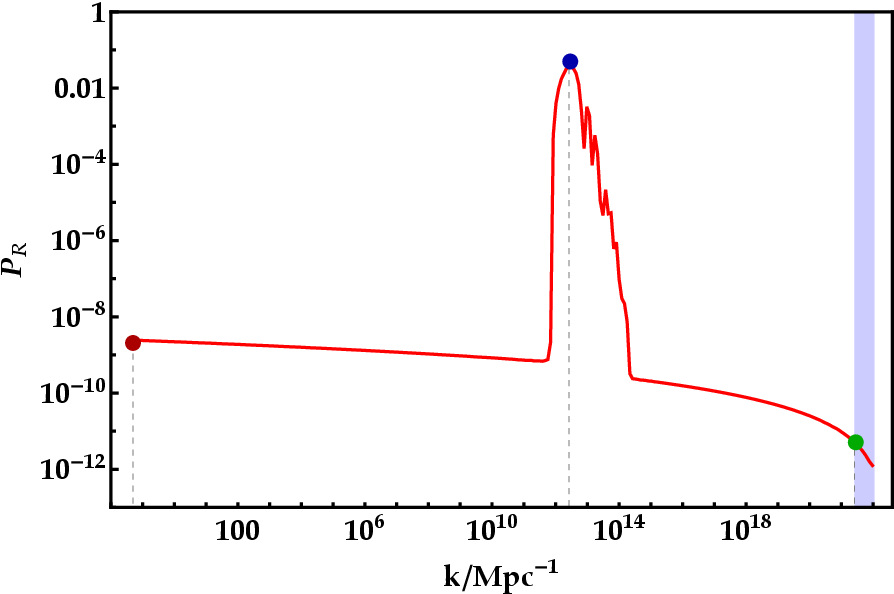}}\hspace{.1cm}
\subfigure{ \includegraphics[width=.468\textwidth]%
{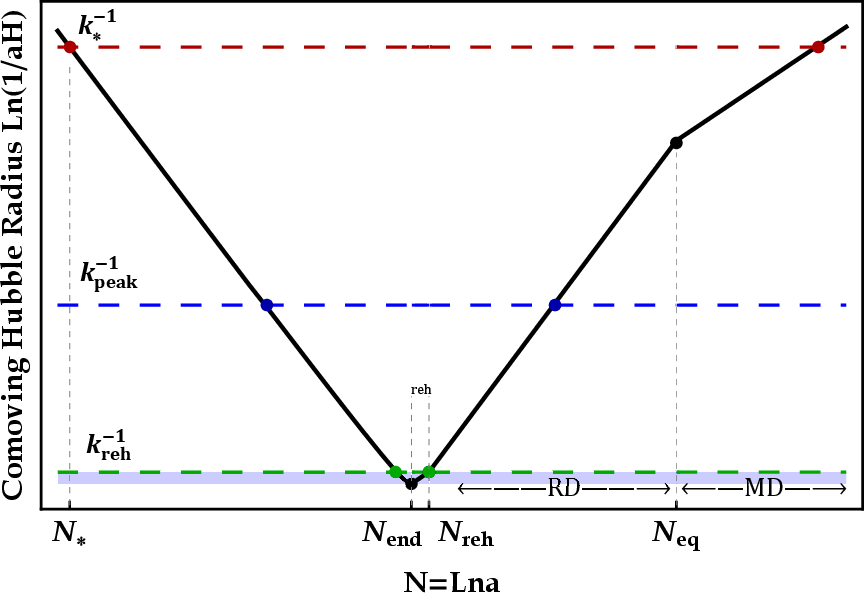}}\hspace{.1cm}
\subfigure{\includegraphics[width=.490\textwidth]%
{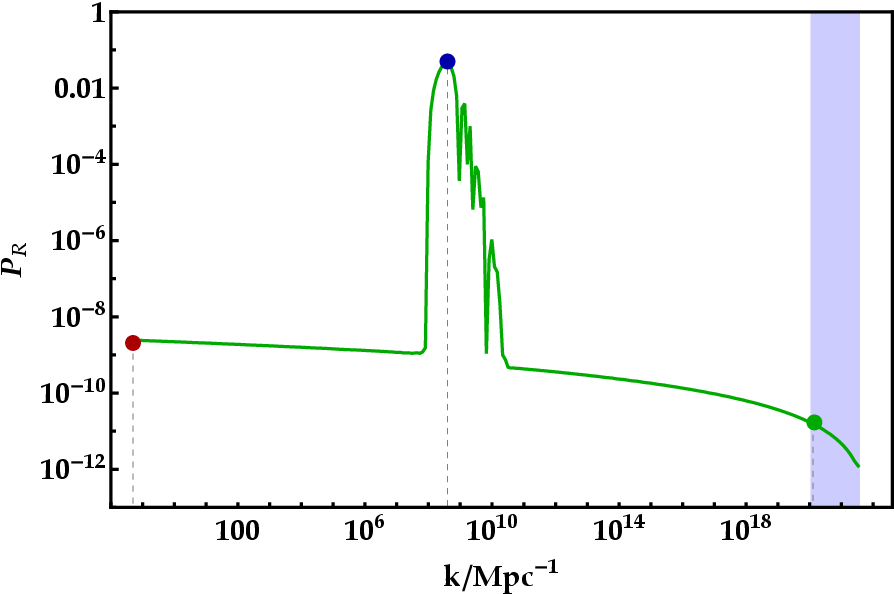}} \hspace{.1cm}
\subfigure{ \includegraphics[width=.468\textwidth]%
{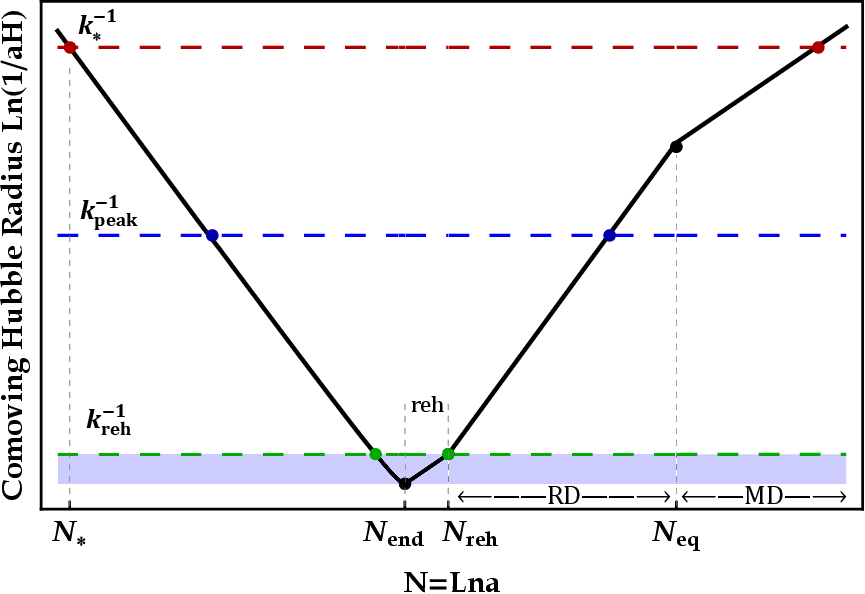}}\hspace{.1cm}
\subfigure{ \includegraphics[width=.490\textwidth]%
{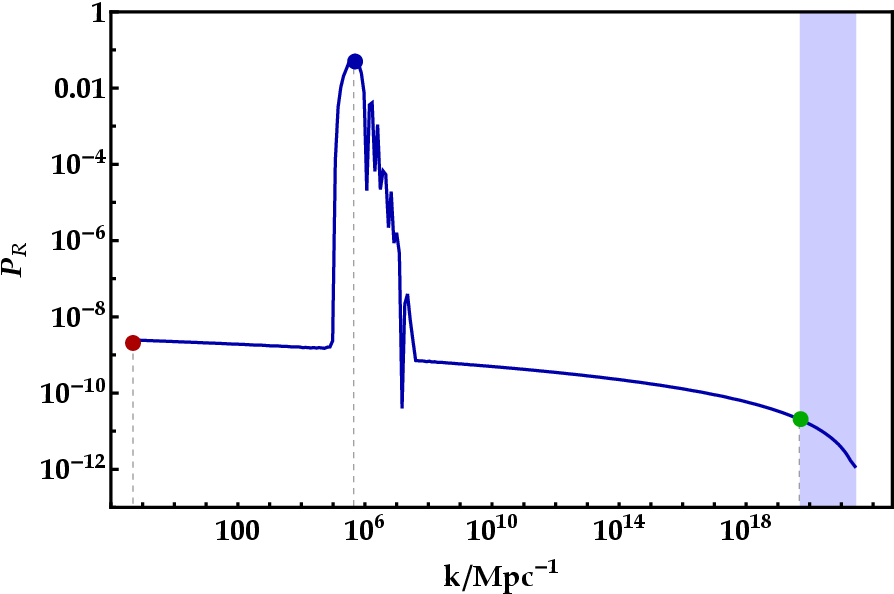}}\hspace{.1cm}
\subfigure{\includegraphics[width=.468\textwidth]%
{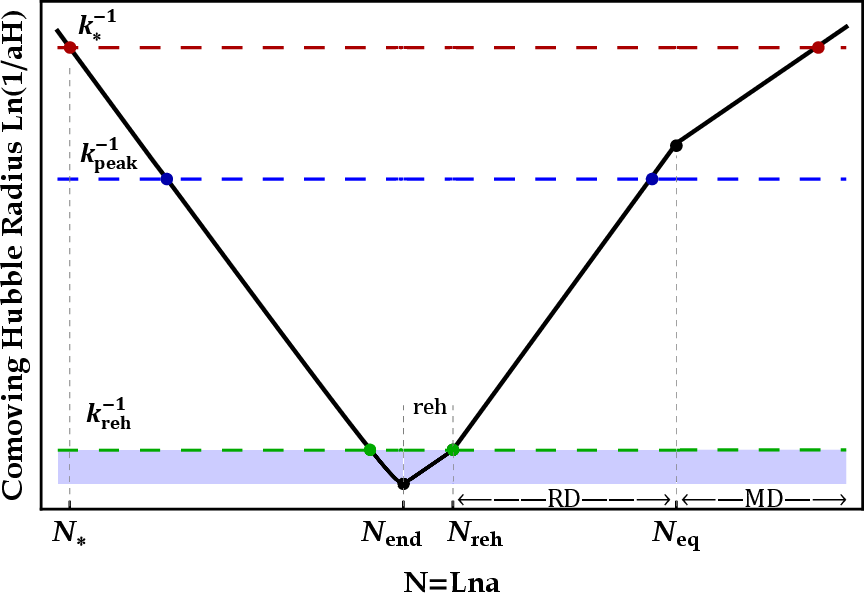}}\hspace{.1cm}
\subfigure{ \includegraphics[width=.490\textwidth]%
{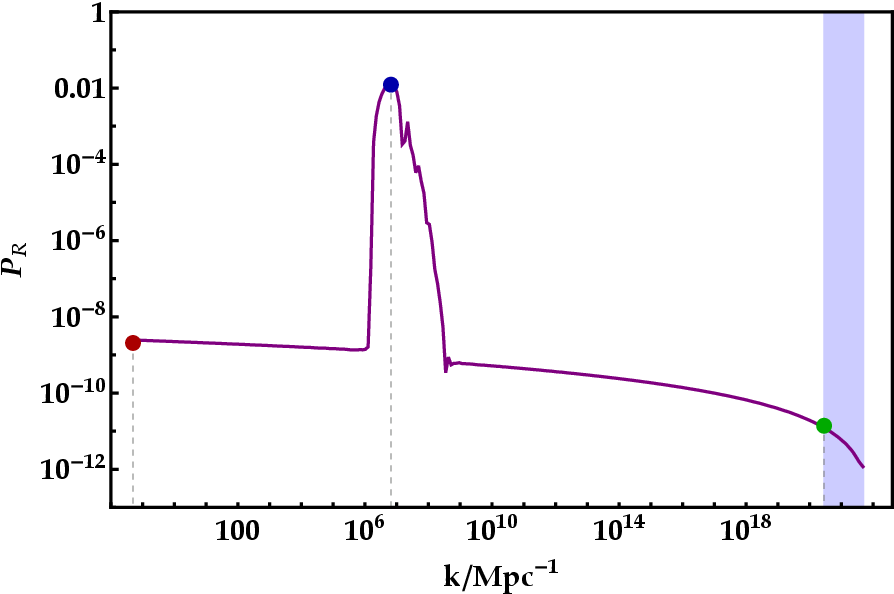}}\hspace{.1cm}
\subfigure{ \includegraphics[width=.468\textwidth]%
{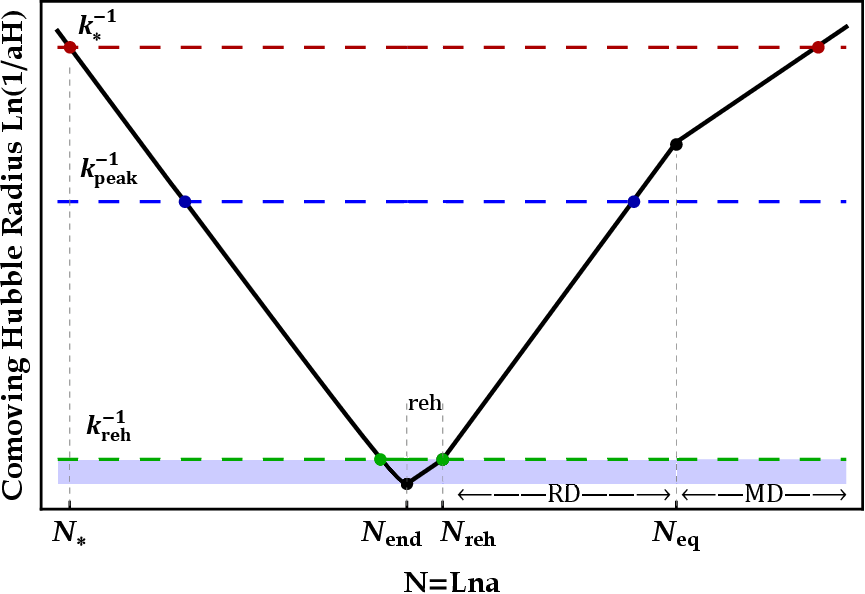}}\hspace{.1cm}
\end{minipage}
\caption{(Left) The curvature power spectra with regard to wavenumber $k$ for cases A (red curve), B (green curve), C (blue curve) and D (purple curve).  Here, the red, blue and green circles depict $k_*$, $k_{\text{peak}}$ and $k_{\text{reh}}$. (Right) The comoving Hubble radius against $N$, in which the red, blue and green circles depict the horizon leaving (reverting) for the scales $k_*^{-1}$, $k_{\text{peak}}^{-1}$ and $k_{\text{reh}}^{-1}$, respectively. The shadowy portions correspond to the reverting scales to the horizon during the reheating stage.}
\label{fig:reh}
\end{figure*}

\begin{figure}
\begin{minipage}[b]{1\textwidth}
\subfigure{ \includegraphics[width=.490\textwidth]%
{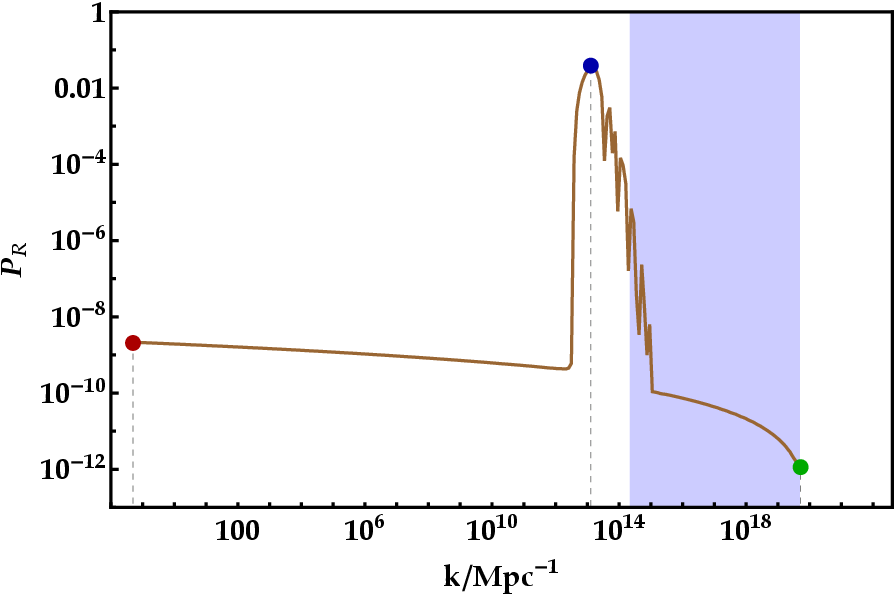}}\hspace{.1cm}
\subfigure{ \includegraphics[width=.468\textwidth]%
{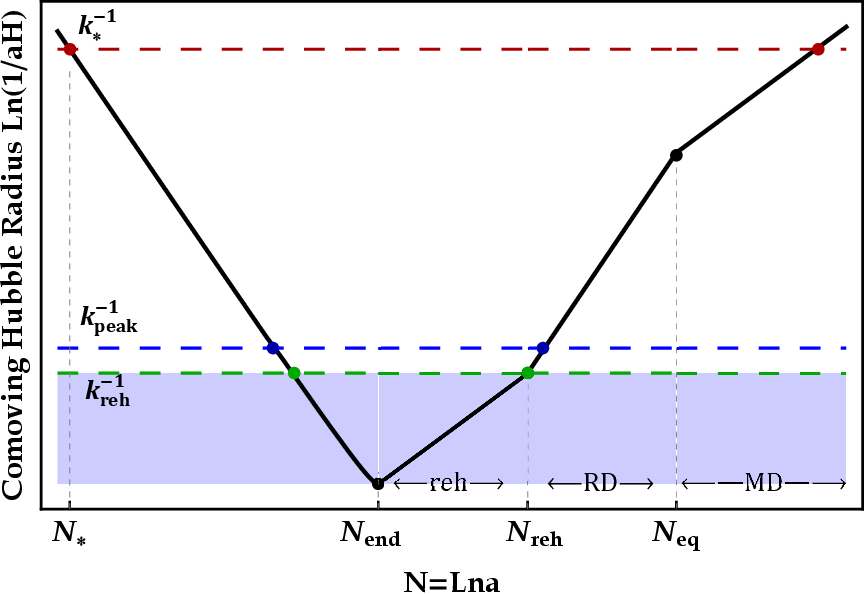}}\hspace{.1cm}
\end{minipage}
\caption{Same as Figs. \ref{fig:reh}, but for case F (brown curve) realizing slow-reheating period with $N_{\text{reh}}=24.63$ and low temperature $T_{\text{reh}}=1.33\times10^{7} {\rm GeV}$.}
\label{fig:reh2}
\end{figure}

In this section, we are interested in analyzing the reheating era in order to know when the super-horizon perturbed scales revet to the horizon after inflation. Among the perturbed scales, those ones which pertinent to peaks of the curvature power spectra (see Table \ref{tab3} and blue circles in the left panels of Figs. \ref{fig:reh}-\ref{fig:reh2}) are important due to their key roles in producing PBHs seeds. A lengthy reheating era with low temperature after inflation  may result in reverting the peak scales to the horizon in the reheating stage rather than the RD era. Hence, analyzing the duration of reheating era ($N_{\text{reh}}$) and its temperature $T_{\rm reh}$ are essential. The duration of the reheating stage $N_{\text{reh}}$ is specified by the $e$-folds number between the end of inflation and end of reheating, which is related to the duration of the inflationary era $\Delta N=N_{\rm end}-N_*$ as follows \citep{Heydari:2022,safaii:2024,Solbi:2024,Dalianis:2019,mahbub:2020}
\begin{eqnarray}
N_{\text{reh}}&=&\frac{4}{1-3\omega_{\rm reh}}\left[61.6-\ln\left (\frac{\rho_{\text{end}}^{1/4}}{H_{\ast}}\right)-\Delta N\right],\label{eq:Nre}
\end{eqnarray}
where, $\omega_{\rm reh}$ is the equation of state parameter of the reheating stage and is taken as $\omega_{\rm reh}=0$ in this model. The energy density in the end of inflation is computed through $\rho_{\text{end}}=3M_P^2H_{\rm end}^2$ and $H_*$ is the Hubble parameter at the horizon leaving $e$-fold number $N_*=0$ by the pivot scale $k_*\sim0.05~ \rm Mpc^{-1}$. Pursuant to  \citep{Heydari:2022,safaii:2024,Solbi:2024,Dalianis:2019,mahbub:2020} and using  $N_{\text{reh}}$, the reheating temperature is calculated as
\begin{eqnarray}
T_{\text{reh}}&=&\left(\frac{30}{\pi^{2}g_{*}}\rho_{\text{end}}\right)^{1/4}\exp\left[{-\frac{3}{4}( 1+\omega_{\rm reh})
N_{\text{reh}}}\right],\label{Treh}
\end{eqnarray}
in which $g_*=106.75$ denotes the effective number of relativistic degree of freedom in the energy density of the reheating stage.  Table \ref{tab4} contains the computed values for $N_{\text{reh}}$ and $T_{\rm reh}$ as to each case of the model. Furthermore, this Table embodies duration of the inflationary era $\Delta N$, the $e$-fold number  $N_{k}^{\text{peak}}$ and wavenumber $k_{\text{peak}}$ as to the peak position of the curvature power spectrum (at $\phi=\phi_c$ corresponding to the blue circles in the left panels of Figs. \ref{fig:reh} and \ref{fig:reh2}) and $k_{\rm reh}$ as well. The $k_{\rm reh}$ indicates the wavenumber of the reverting scale to the horizon in the end of reheating era and is computed as \citep{Dalianis:2019}
\begin{equation}\label{kreh}
 k_{\text{reh}}=e^{-N_{\text{reh}}/2}k_{\text{end}},
\end{equation}
wherein, $k_{\text{end}}$ is the wavenumber of the scale that exits from (reverts to) the horizon at the end of inflation.

In what follows, we attempt to show that in our model the scales pertinent to PBHs seeds (blue circles in the left panels of Figs. \ref{fig:reh} and \ref{fig:reh2}) revert to the horizon in RD era after reheating stage. To this end, the scale $k_{\text{reh}}^{-1}$ is calculated for each case of the model and compared to the peak scale $k_{\text{peak}}^{-1}$ of that case. It is known that, the larger scales with the larger $k^{-1}$ leave the horizon earlier and thereafter revert to the horizon later than smaller scales with the smaller $k^{-1}$. As regards $k_{\text{reh}}^{-1}$ corresponds to the scale which reverts to the horizon at the end of reheating, thus each scale larger than  $k_{\text{reh}}^{-1}$ revert to the horizon after the reheating era in RD or MD era. Therefore, drawing a comparison between the listed values of $k_{\text{reh}}$ and  $k_{\text{peak}}$  in Table \ref{tab4} illustrates that the peak scale for each case of the model is larger than $k_{\text{reh}}^{-1}$. As a result, the peak scales for all cases of our model revert to the horizon in the RD era after the reheating to produce PBHs.

The above outcomes are exhibited in Figs. \ref{fig:reh} and \ref{fig:reh2} as well. In these figures, the curvature power spectra versus $k$ in the left panels and comoving Hubble radius versus $N$ in the right panels are schemed for all cases of the model. Altogether in Figs. \ref{fig:reh}  and \ref{fig:reh2}, (i) the shadowy portions in all panels indicate the scales which re-enter the horizon during the reheating stage. (ii) The red, blue and green circles on the curvature power spectra in the left panels depict, respectively, the pivot scale $k_*\sim0.05~ \rm Mpc^{-1}$, the peak scale $k_{\text{peak}}$ and the reverting scale to the horizon in the end of reheating era $k_{\text{reh}}$. (iii) The red, blue and green circles in right panels indicate exit from (revert to) the horizon as to the scales  $k_*^{-1}$, $k_{\text{peak}}^{-1}$ and $k_{\text{reh}}^{-1}$, respectively. Consequently, the PBHs scales in our model revert to the horizon in RD era to produce ultra-dense portions to collapse and born PBHs.

\begin{table*}[ht!]
  \centering
  \caption{The quantities of $\Delta N$, $N_{\text{reh}}$, $N_{k}^{\text{peak}}$, $k_{\text{peak}}$, $k_{\rm reh}$ and $T_{\text{reh}}$ as to all cases of the model.}
\begin{tabular}{ccccccc}
  \hline
   $\#$ & \quad$\Delta N$\quad &\quad$N_{\text{reh}}$\quad &\quad $N_{k}^{\text{peak}}$\quad &\quad$k_{\text{peak}}/\text{Mpc}^{-1}$\quad& \quad$k_{\rm reh}/\rm Mpc^{-1}$\quad& \quad$T_{\text{reh}}/\rm GeV$\\ \hline\hline
  Case A &\quad56.32\quad  &\quad 2.90\quad& \quad32.2\quad & \quad$3.06\times10^{12}$ \quad&\quad $2.56\times10^{21}$\quad &\quad$1.56\times10^{14}$\\ \hline
 Case B &\quad  55.26\quad  & \quad  7.13\quad &\quad23.0\quad&\quad  $3.59\times10^{8}$ \quad&\quad$1.07\times10^{20}$\quad &\quad $6.56\times10^{12}$\\ \hline
 Case C &\quad55.01\quad  & \quad8.15\quad & \quad16.0\quad & \quad$3.63\times10^{5}$\quad &\quad$4.98\times10^{19}$\quad &\quad$3.04\times10^{12}$\\ \hline
 Case D &\quad55.57\quad  & \quad5.90\quad & \quad19.0\quad & \quad$6.90\times10^{6}$\quad &\quad$2.69\times10^{20}$\quad &\quad$1.65\times10^{13}$\\ \hline
  Case F &\quad$50.85$\quad  & \quad$24.63$\quad & \quad$33.8$\quad & \quad$1.36\times10^{13}$\quad &\quad$2.18\times10^{14}$\quad &\quad$1.33\times10^{7}$\\ \hline
\end{tabular}
\label{tab4}
\end{table*}
It is known that the duration of reheating era is only constrained by the requirement that the Universe must have been RD during BBN. It means that, reheating could have continued up to the energy scale of BBN, $T_{\text{reh}}=10~\text{MeV}$ \citep{German:2023}. The small reheating durations obtained for cases A to D of the model with temperatures around the inflation scale (see Table \ref{tab4}) are due to the maximum values of the duration of inflationary era around $56$ $e$-folds. Since, we are interested in cases with $r-n_s$ spot being inside the 68$\%$ CL of the Planck diagram. Hence, for the lower reheating temperature around $T_{\text{reh}}=10^7~\text{GeV}$ in case F, the duration of inflation could decrease to around $\Delta N=50.85$ and the slow-reheating era could take place with $N_{\text{reh}}=24.63$ (as shown in Table \ref{tab4}). However, the $r-n_s$ spot for case F remains in the $68\%$ CL of the Planck diagram  (see Table \ref{tab3}). It is obvious from the panels of Fig. \ref{fig:reh2} and the last row of Table \ref{tab4} that, similar to other cases of the model, the peak scale of the curvature power spectrum for case F returns to the horizon after the reheating stage in the RD era too.

Note that for the lower reheating temperature, the duration of reheating increases while the inflationary $e$-folds number decreases. Ergo, the peak scale of the scalar power spectrum may return to the horizon in the reheating era to form PBHs.  The criteria for PBHs formation and GWs generation should be modified if the peak of the primordial power spectrum reenters the cosmological horizon during the reheating stage. In \citep{Dalianis:2019}, it is illustrated that, PBHs could be generated during the reheating era from the smaller peak in the scalar power spectrum around ${\cal O}(10^{-5})$. The density threshold in the reheating era is smaller than the RD, ergo the PBHs could be formed from the smaller amplitude of the scalar perturbations. Generation of PBHs and GWs in the reheating era are not in the scope of this study and we leave them to  upcoming works (for further information on this issue see \citep{Dalianis:2019,Padilla:2024,Padilla:2024b}). In the following sections, we are interested in calculating PBHs and GWs for cases of Tables \ref{tab1}-\ref{tab2}, in the RD era.

\section{PBHs abundance}\label{sec5}
\begin{figure*}
\centering
\includegraphics[scale=0.7]{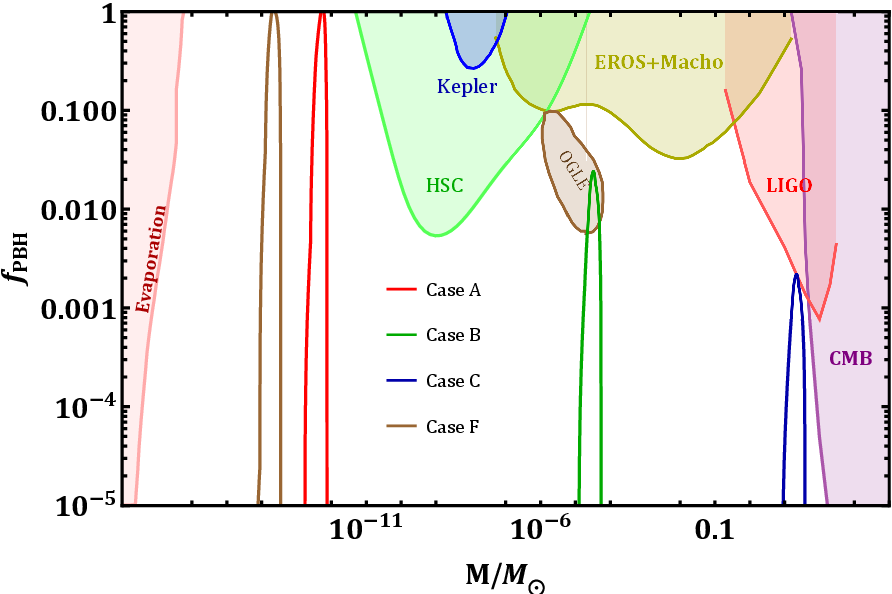}
\caption{The PBHs mass spectra concerning cases A (red curve), B (green curve), C (blue curve) and F (brown curve). The observational restricted portions are demonstrated by colory domains. The disallowed portions are prohibited by observations of CMB \citep{CMB} (purple portion), LIGO-VIRGO event \citep{Abbott:2019,Chen:2020,Boehm:2021,Kavanagh:2018} (red portion), microlensing events through MACHO \citep{MACHO} and EROS \citep{EORS} (khaki portion), Kepler \citep{Kepler} (blue portion), Icarus \citep{Icarus}, OGLE \citep{OGLE-1,OGLE-2} and Subaru-HSC \citep{subaro} (green portion), and PBHs evaporation \citep{EGG, Laha:2019,Clark,Shikhar:2022,Dasgupta} (pink portion). The only permissible portion (brown portion) in the plot is defined by the ultrashort-timescale microlensing events in the OGLE data \citep{OGLE-1,OGLE-2}.}
\label{fig-fpbh}
\end{figure*}

In the erstwhile section, the curvature power spectra for cases of the GPLNC natural inflationary model were computed through the FR mechanism.
Here, using the computed values for ${\cal P}_{\cal R}$, the  abundances $f_{\rm{PBH}}$ and masses $M_{\rm{PBH}}$ of resulted PBHs are evaluated. The PBHs abundance $f_{\rm{PBH}}\equiv\Omega_{\rm {PBH}}/\Omega_{\rm{DM}}$ specifies that, what percentage of the current DM budget of the cosmos is allotted to PBHs. The current density parameter of DM is defined by Planck 2018 data as $\Omega_{\rm {DM}}h^2\simeq0.12$ \citep{akrami:2018}. The fractional PBHs abundance is given by
\begin{equation}\label{fPBH}
f_{\rm{PBH}}(M)\equiv\frac{\Omega_{\rm {PBH}}}{\Omega_{\rm{DM}}}= \frac{\beta(M)}{1.84\times10^{-8}}\left(\frac{\gamma}{0.2}\right)^{3/2}\left(\frac{g_*}{10.75}\right)^{-1/4}
\left(\frac{0.12}{\Omega_{\rm{DM}}h^2}\right)
\left(\frac{M}{M_{\odot}}\right)^{-1/2},
\end{equation}
where, $\gamma=\big(1/\sqrt{3}\big)^{3}$ signifies the collapse efficiency parameter \citep{carr:1975} and  $g_{*}=106.75$ is the effective number of relativistic degrees of freedom. Moreover,  considering the gaussian distribution for the primordial perturbations, the production rate of PBHs $\beta(M)$ is computed through the Press-Schechter formulation as
 \cite{Tada:2019,young:2014}
\begin{equation}\label{betta}
  \beta(M)=\int_{\delta_{c}}^{\infty}\frac{{\rm d}\delta}{\sqrt{2\pi\sigma^{2}(M)}}e^{-\frac{\delta^{2}}{2\sigma^{2}(M)}}=\frac{1}{2}~ {\rm erfc}\left(\frac{\delta_{c}}{\sqrt{2\sigma^{2}(M)}}\right),
\end{equation}
where "erfc" is the error function complementary and $\delta_{c}=0.4$ is the density threshold  \citep{Musco:2013,Harada:2013}. Additionally,  $\sigma^{2}(M)$ is the smoothed density contrast on the scale $k$ and  is related to the curvature power spectrum ${\cal P}_{\cal R}$ and  Gaussian window function $W(x)=\exp{\left(-x^{2}/2 \right)}$ as follows
\begin{equation}\label{sigma}
\sigma_{k}^{2}=\left(\frac{16}{81} \right) \int_0^{\infty}\frac{{\rm d}q}{q} W^{2}(q/k)(q/k)^{4} {\cal P}_{\cal R}(q).
\end{equation}
At last $M$ in Eq. (\ref{fPBH}) expresses the PBHs mass and is associated to the horizon mass as
\begin{align}\label{Mpbheq}
M(k)\equiv\gamma M_{\text{horizon}}\simeq M_{\odot} \left(\frac{\gamma}{0.2} \right) \left(\frac{10.75}{g_{*}} \right)^{\frac{1}{6}} \left(\frac{k}{1.9\times 10^{6}\rm Mpc^{-1}} \right)^{-2}.
\end{align}
Thereafter,  we are able to calculate the abundances $f_{\rm PBH}$ and masses $M_{\rm PBH}$ of PBHs as to cases of Table \ref{tab2}, using the obtained ${\cal P}_{\cal R}$ and  Eqs. (\ref{fPBH})-(\ref{Mpbheq}). The numerical results for $f_{\rm PBH}^{\text{peak}}$ and $M_{\text{PBH}}^{\text{peak}}$ (corresponding to the peak spot $\phi=\phi_c$) are sorted into Table \ref{tab3}.  Into the bargain, in Fig. \ref{fig-fpbh} the evaluated results for $f_{\rm PBH}$ with regard to  $M_{\rm PBH}$  in addition to observational restricted portions are plotted for all cases. This figure illustrates that, (i) the corresponding PBHs to cases A (red curve) and F (brown curve) with $f_{\rm PBH}^{\text{peak}}\sim 1$ could comprise all DM budget of the Universe. (ii) The mass spectrum of PBH as to case B (green curve)  lies in the allowed data of OGLE \citep{OGLE-1,OGLE-2}, so it could be a considerable source for ultrashort-timescale microlensing events. (iii)  The  current GWs spectrum  propagated from PBH of the case C (blue curve) could be tracked down by LIGO and Virgo interferometers, hence this case of PBHs could be a considerable origin for LIGO-Virgo events. It is notable that, the abundance of PBH of the case D is of order $10^{-27}$ and it has a trivial contribution in the DM budget.
\section{ secondary gravitational waves}\label{sec6}
Concurrent with generation of the seeds of PBHs in the RD era, secondary GWs propagate in the Universe. The density spectrum of these GWs could be traced via various observatories such as SKA \citep{ska}, EPTA  \citep{EPTA-a,EPTA-b,EPTA-c}, LISA  \citep{lisa,lisa-a},  BBO  \citep{Yagi:2011} and DECIGO  \citep{Yagi:2011,Seto:2001} at present. Thus, the PBHs with different masses could be detected by way of the propagated secondary GWs, indirectly. In this section, the density spectra of the secondary GWs of the generated PBHs in the natural GPLNC inflationary model are evaluated. the present-time GWs density parameter $\Omega_{\rm GW_0}$ at $\eta_0$, is associated with its match at time $\eta$ as follows ($\eta$ signifies the conformal time) \citep{Inomata:2019-a}
\begin{eqnarray}\label{OGW0}
\Omega_{\rm GW_0}h^2 = 0.83\left( \frac{g_{*}}{10.75} \right)^{-1/3}\Omega_{\rm r_0}h^2\Omega_{\rm{GW}}(\eta,k),
\end{eqnarray}
therein, $g_{*}\simeq106.75$ expresses the effective degrees of freedom in the energy density at $\eta$ and  $\Omega_{\rm r_0}h^2\simeq 4.2\times 10^{-5}$ is the present-time density parameter of radiation. Moreover,  $\Omega_{\rm{GW}}(\eta,k)$ is calculated via \citep{Kohri:2018,Lu:2019}
\begin{equation}
 \label{OmegaGW}
 \Omega_{\rm GW}(k,\eta)=\frac{1}{6}\left(\frac{k}{aH}\right)^{2}\int_{0}^{\infty}dv\int_{|1-v|}^{|1+v|}du\left(\frac{4v^{2}-\left(1-u^{2}+v^{2}\right)^{2}}{4uv}\right)^{2}\overline{I_{RD}^{2}(u,v,x)}\mathcal{P}_{\cal R}(ku)\mathcal{P}_{\cal R}(kv),
\end{equation}
where, the time-mean of the source term is expressed by
\begin{align}
 \overline{I_{RD}^{2}(u,v,x\to\infty)}= & \frac{1}{2x^{2}}\Bigg[\left(\frac{3\pi\left(u^{2}+v^{2}-3\right)^{2}\Theta\left(u+v-\sqrt{3}\right)}{4u^{3}v^{3}}+\frac{T_{c}(u,v,1)}{9}\right)^{2}
 \nonumber\\
 & +\left(\frac{\tilde{T}_{s}(u,v,1)}{9}\right)^{2}\Bigg].
 \label{IRD2b}
\end{align}
In the above equation, $\Theta$ expresses the Heaviside step function where, $\Theta(x)=0$ for $x<0$ and $\Theta(x)=1$ for $x\geq 0$. Furthermore, $T_c$ and $T_s$ are specified as follows
\begin{align}
T_{c}= & -\frac{27}{8u^{3}v^{3}x^{4}}\Bigg\{-48uvx^{2}\cos\left(\frac{ux}{\sqrt{3}}\right)\cos\left(\frac{vx}{\sqrt{3}}\right)\left(3\sin(x)+x\cos(x)\right)+
\nonumber\\
& 48\sqrt{3}x^{2}\cos(x)\left(v\sin\left(\frac{ux}{\sqrt{3}}\right)\cos\left(\frac{vx}{\sqrt{3}}\right)+u\cos\left(\frac{ux}{\sqrt{3}}\right)\sin\left(\frac{vx}{\sqrt{3}}\right)\right)+
\nonumber\\
& 8\sqrt{3}x\sin(x)\Bigg[v\left(18-x^{2}\left(u^{2}-v^{2}+3\right)\right)\sin\left(\frac{ux}{\sqrt{3}}\right)\cos\left(\frac{vx}{\sqrt{3}}\right)+
\nonumber\\
& u\left(18-x^{2}\left(-u^{2}+v^{2}+3\right)\right)\cos\left(\frac{ux}{\sqrt{3}}\right)\sin\left(\frac{vx}{\sqrt{3}}\right)\Bigg]+
\nonumber\\
& 24x\cos(x)\left(x^{2}\left(-u^{2}-v^{2}+3\right)-6\right)\sin\left(\frac{ux}{\sqrt{3}}\right)\sin\left(\frac{vx}{\sqrt{3}}\right)+
\nonumber\\
& 24\sin(x)\left(x^{2}\left(u^{2}+v^{2}+3\right)-18\right)\sin\left(\frac{ux}{\sqrt{3}}\right)\sin\left(\frac{vx}{\sqrt{3}}\right)\Bigg\}
\nonumber\\
& -\frac{\left(27\left(u^{2}+v^{2}-3\right)^{2}\right)}{4u^{3}v^{3}}\Bigg\{\text{Si}\left[\left(\frac{u-v}{\sqrt{3}}+1\right)x\right]-\text{Si}\left[\left(\frac{u+v}{\sqrt{3}}+1\right)x\right]
\nonumber\\
& +\text{Si}\left[\left(1-\frac{u-v}{\sqrt{3}}\right)x\right]-\text{Si}\left[\left(1-\frac{u+v}{\sqrt{3}}\right)x\right]\Bigg\},
\label{Tc}
\end{align}

\begin{align}
T_{s}= & \frac{27}{8u^{3}v^{3}x^{4}}\Bigg\{48uvx^{2}\cos\left(\frac{ux}{\sqrt{3}}\right)\cos\left(\frac{vx}{\sqrt{3}}\right)\left(x\sin(x)-3\cos(x)\right)-
\nonumber\\
& 48\sqrt{3}x^{2}\sin(x)\left(v\sin\left(\frac{ux}{\sqrt{3}}\right)\cos\left(\frac{vx}{\sqrt{3}}\right)+u\cos\left(\frac{ux}{\sqrt{3}}\right)\sin\left(\frac{vx}{\sqrt{3}}\right)\right)+
\nonumber\\
& 8\sqrt{3}x\cos(x)\Bigg[v\left(18-x^{2}\left(u^{2}-v^{2}+3\right)\right)\sin\left(\frac{ux}{\sqrt{3}}\right)\cos\left(\frac{vx}{\sqrt{3}}\right)+
\nonumber\\
& u\left(18-x^{2}\left(-u^{2}+v^{2}+3\right)\right)\cos\left(\frac{ux}{\sqrt{3}}\right)\sin\left(\frac{vx}{\sqrt{3}}\right)\Bigg]+
\nonumber\\
& 24x\sin(x)\left(6-x^{2}\left(-u^{2}-v^{2}+3\right)\right)\sin\left(\frac{ux}{\sqrt{3}}\right)\sin\left(\frac{vx}{\sqrt{3}}\right)+
\nonumber\\
& 24\cos(x)\left(x^{2}\left(u^{2}+v^{2}+3\right)-18\right)\sin\left(\frac{ux}{\sqrt{3}}\right)\sin\left(\frac{vx}{\sqrt{3}}\right)\Bigg\}-\frac{27\left(u^{2}+v^{2}-3\right)}{u^{2}v^{2}}+
\nonumber\\
& \frac{\left(27\left(u^{2}+v^{2}-3\right)^{2}\right)}{4u^{3}v^{3}}\Bigg\{-\text{Ci}\left[\left|1-\frac{u+v}{\sqrt{3}}\right|x\right]+\ln\left|\frac{3-(u+v)^{2}}{3-(u-v)^{2}}\right|+
\nonumber\\
& \text{Ci}\left[\left(\frac{u-v}{\sqrt{3}}+1\right)x\right]-\text{Ci}\left[\left(\frac{u+v}{\sqrt{3}}+1\right)x\right]+\text{Ci}\left[\left(1-\frac{u-v}{\sqrt{3}}\right)x\right]\Bigg\},
\label{Ts}
\end{align}
where, the sine-integral $\text{Si}(x)$ and cosine-integral $\text{Ci}(x)$ functions are designated via
\begin{equation}
 \label{SiCi}
 \text{Si}(x)=\int_{0}^{x}\frac{\sin(y)}{y}dy,\qquad\text{Ci}(x)=-\int_{x}^{\infty}\frac{\cos(y)}{y}dy.
\end{equation}
Additionally, the function $\tilde{T}_{s}(u,v,1)$ is specified as
\begin{equation}
 \label{Tst}
 \tilde{T}_{s}(u,v,1)=T_{s}(u,v,1)+\frac{27\left(u^{2}+v^{2}-3\right)}{u^{2}v^{2}}-\frac{27\left(u^{2}+v^{2}-3\right)^{2}}{4u^{3}v^{3}}\ln\left|\frac{3-(u+v)^{2}}{3-(u-v)^{2}}\right|.
\end{equation}
In the end, the wavenumber can be changed to frequency from $f=1.546 \times 10^{-15}\left( k/{\rm Mpc}^{-1}\right){\rm Hz}$.

\begin{figure*}
\centering
\includegraphics[scale=0.6]{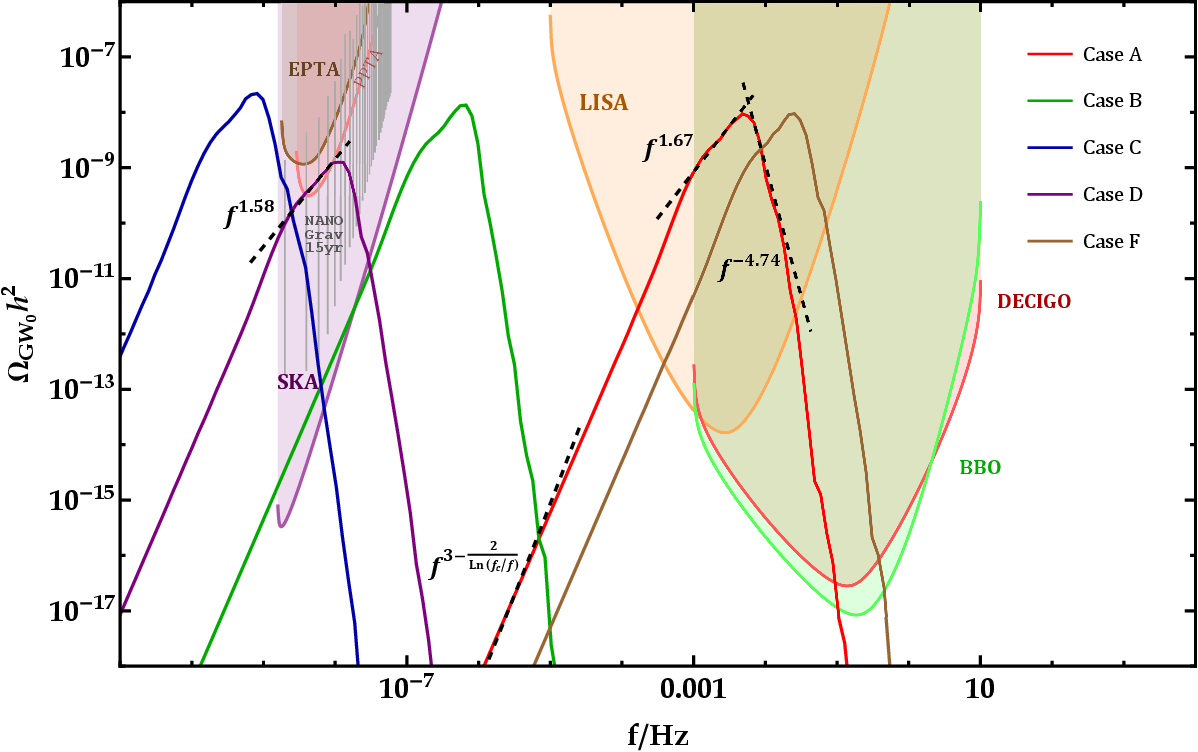}
\vspace{-0.5em}
\caption{The  present-time density spectra of the secondary gravitational waves  $\Omega_{\rm GW_0}h^2$ with regard to frequency for cases A (red curve), B (green curve), C (blue curve), D (Purple curve) and F (brown curve) as well as the observational dominions of EPTA (brown domain), SKA (purple domain), NANOGrav 15 year data (gray shadowy area) LISA (orange domain), DECIGO (red domain) and BBO (green domain).  The  power-like functioning of $\Omega_{\rm GW_0}h^2$ spectra have been represented by dashed black lines in various frequency ranges according to Table \ref{table:GWs} for cases A and D.}
\label{fig-omega}
\end{figure*}

In this stage, we are able to calculate the present-time density spectra of secondary GWs relevant to all cases of the model via the numerical values of  ${\cal P}_{\cal R}$  and Eqs. (\ref{OGW0})-(\ref{Tst}). The numerical outcomes for $\Omega_{\rm GW_0}$ with regard to frequency have been illustrated with diagram of Fig. \ref{fig-omega} for cases A (red curve), B (green curve), C (blue curve), D(purple curve) and F (brown curve). In addition, in this figure the sensibility dominions of GWs observatories are shown by purple (SKA), brown (EPTA), orange (LISA), green  (BBO) and red (DECIGO) areas. Moreover, the shadowy gray area in this Figure represents the dominion of the NANOGrav 15 year data.
It is inferred from Fig. \ref{fig-omega} that, (i) the $\Omega_{\rm GW_0}$ spectra related to cases A and F can be detected in light of the oncoming data of LISA, DICIEGO and BBO observatories. (ii) The spectra of $\Omega_{\rm GW_0}$ related to cases B, C and D could be ferreted out through the SKA detector. (iii) The peak of $\Omega_{\rm GW_0}$ for case D lies inside the NANOGrav 15 year and PPTA data, hence this case can be contemplated as a promising origin for the low-frequency GWs signals found by  NANOGrav, EPTA, PPTA and CPTA \citep{NANOGrav-1,NANOGrav-2,NANOGrav-3,NANOGrav-4,NANOGrav-5,EPTA-1,EPTA-2,EPTA-3,
EPTA-4,EPTA-5}. (iv) The $\Omega_{\rm GW_0}$ spectra show power-like functioning with regard to the frequency $f$ in various frequency ranges (see the black dashed line fitted on the $\Omega_{\rm GW_0}$ spectra of cases A and D).

Into the bargain, the frequencies of the peaks of $\Omega_{\rm GW_0}$ spectra ($f_c$) as well as the peak values for $\Omega_{\rm GW_0}$ for all cases of the model are arranged in Table \ref{table:GWs}.
It is proven that, the $\Omega_{\rm GW_0}$ spectrum exhibits the power-like functioning versus the frequency as $\Omega_{\rm GW_0} (f) \sim f^{n} $ \citep{fu:2020,Xu,Kuroyanagi}. The evaluated exponents $n$ in frequency scopes $f\ll f_{c}$,  $f<f_{c}$ and  $f>f_{c}$ for all cases of the model have been registered in Table \ref{table:GWs}. It is deduced from the sorted data in this table that, the evaluated exponents $n$ in the infrared frequency scope $f\ll f_{c}$ for all cases of the model satisfy the logarithmic relation $n=3-2/\ln(f_c/f)$ analyzed in \citep{Yuan:2020,shipi:2020,Yuan:2023}. Furthermore, the exponent $n$ in the NANOGrav dominion of the frequency scope $f<f_{c}$ is related to the spectral index  $\gamma$ as $n=5-\gamma$. The imposed constraint on  the $\gamma$ parameter by the NANOGrav collaboration is $\gamma=3.2\pm0.6$ \citep{NANOGrav-1,NANOGrav-2,NANOGrav-3,NANOGrav-4,NANOGrav-5}. Ergo, the exponent $n=1.58$ for case D of our model results in $\gamma= 3.42$ which satisfies the NANOGrav constraint (see Table \ref{table:GWs} and Fig. \ref{fig-omega}).
\begin{table*}[ht!]
  \centering
  \caption{The peak values of frequencies and $\Omega_{\rm GW_0}h^2$ spectra, in addition to the exponent $n$ in frequency ranges $f\ll f_{c}$,  $f<f_{c}$ and  $f>f_{c}$ in regard to cases A, B, C, D and F.}
\scalebox{1}[1] {
\begin{tabular}{cccccc}
\hline
\#  & $\qquad\qquad$ $f_{c}/{\rm Hz}$ $\qquad\qquad$ & $\quad$ $\Omega_{\rm GW_0}h^2\left(f_{c}\right)$ $\quad$ & $\quad$ $n_{f\ll f_{c}}$ $\quad$ & $\quad$ $n_{f<f_{c}}$ $\quad$ & $\quad$ $n_{f>f_{c}}$\tabularnewline
\hline
\hline
Case A & $4.74\times10^{-3}$ & $9.37\times10^{-9}$ & $3.05$ & $1.67$ & $-4.74$ \tabularnewline
\hline
Case B & $6.68\times10^{-7}$ & $1.35\times10^{-8}$ & $3.09$ & $1.57$ & $-1.65$\tabularnewline
\hline
Case C & $8.12\times10^{-10}$ & $2.18\times10^{-8}$ & $3.06$ & $1.67$ & $-1.81$\tabularnewline
\hline
Case D & $1.07\times10^{-8}$ & $1.26\times10^{-9}$ & $3.07$ & $1.58$ & $-1.67$\tabularnewline
\hline
Case F & $2.52\times10^{-2}$ & $9.42\times10^{-9}$ & $3.01$ & $1.60$ & $-6.56$\tabularnewline
\hline
\end{tabular}
    }
  \label{table:GWs}
\end{table*}
\section{Conclusions}\label{sec7}
In this paper, production of PBHs and GWs in the GPLNC natural inflationary model through FR mechanism were probed. In this model, employing the non-canonical mass scale parameter $M(\phi)$ of the power-law Lagrangian (\ref{Lagrangian}) as a twofold function of the scalar field (\ref{eq:M})-(\ref{eq:bump}), a cliff-like region was produced in the field evolution path. When the scalar field fall off the cliff, its kinetic energy increases and a transient FR stage is produced  which lasts for about one or even less than one $e$-fold (see Figs. \ref{phi} and \ref{dphi}). While the non-adiabatic transition takes place in the field evolution from the SR to the FR stage, the particle production results in enhancing the scalar perturbations in the FR stage. Thereafter, horizon reverting of the enhanced scale during the RD era gives rise to the production of PBH seeds. The parameter $M_0$ in $M(\phi)$ and non-canonical $\alpha$ parameter are responsible for reviving the natural potential on the CMB scale in this model. On the other side, the other parameters of $M(\phi)$ like $\{\omega,\phi_c,d\}$ are responsible for producing the suitable cliff-like region in ($\phi=\phi_c$) for accelerating the scalar field in the FR stage to produce PBHs. These three parameters do not affect the CMB scale. During the FR stage, the $\varepsilon_1$ encounters a peaked enhancement, although it remains smaller than one and loyal to the SR condition all over the inflationary era. On the contrary, the $\varepsilon_2$ breaks the SR condition and increases to larger values than one during the FR stage (see Figs. \ref{e1} and \ref{e2}). The main consequences of the chosen model are as follows
\begin{itemize}
  \item The observational predictions of the natural potential in this framework satisfy  the constraints of Planck and BICEP/Keck 2018 (TT,TE,EE+lowE+lensing+BK18+BAO) at 68\%  CL for the scalar spectral index $n_s$ and 95\%  CL for the tensor-to-scalar ratio $r$  \citep{akrami:2018,BK18:2021} (note Table \ref{tab3}).
  \item Two theoretical conjectures of the swampland criteria were established for this model (see plots of Fig. \ref{fig:s}).

  \item Solving the MS equations (\ref{eq:MS}) and (\ref{eq:tMS}) result, respectively, in obtaining the curvature and tensor power spectra for all cases of the model. The curvature spectra ${\cal P}_{\cal R}$ for all cases of the model on the CMB scale $k_*\sim0.05~ \rm Mpc^{-1}$ are of order $10^{-9}$ in consistency with Planck 2018 observations \citep{akrami:2018}, whereas on the smaller scales near $\phi=\phi_c$ (in the FR stage) have peaks of order $10^{-2}$ which are required to produce PBHs (see Fig. \ref{fig:ps}). Also the results of tensor-to-scalar ratio $r={\cal P}_t(k_{*})/{\cal P}_{\cal R}(k_{*})$ for aforementioned model are well compatible with Planck and BICEP/Keck 2018 data.

  \item Through the reheating considerations, it was proven that the peak scales related to PBHs formation revert to the horizon in the RD era to produce PBHs (see Fig. \ref{fig:reh}).
  \item Four cases of PBHs with different masses  were predicted by this model for explaining the DM budget, microlensing events in OGLE data and LIGO/Virgo events (see Table \ref{tab3} and Fig. \ref{fig-fpbh}).
  \item The present-time density spectra  $\Omega_{\rm GW_0}$ of secondary GWs predicted by the model could intersect the sensitivity curves of various GWs observatories in the various frequency ranges (see Fig. \ref{fig-omega}).
  \item The power-like functioning as  $\Omega_{\rm GW_0} (f) \sim f^{n} $ in different frequency zones for all cases of the model were indicated (see Table \ref{table:GWs} and Fig. \ref{fig-omega}).
  \item  In the infrared regime $f\ll f_{c}$, the exponent $n$ satisfied the logarithmic relation $n=3-2/\ln(f_c/f)$ (see Table \ref{table:GWs} and Fig. \ref{fig-omega}).
  \item  The present-time density spectrum of the secondary GWs for case D was located in the NANOGrav 15 year and PPTA data domain (see Fig. \ref{fig-omega}), and it could be a suitable source for the low-frequency GWs detected by NANOGrav, EPTA and PPTA teamwork.
  \item Regarding the relation $n=5-\gamma$ in the frequency scope $f<f_{c}$ of the NANOGrav domain,  the evaluated exponent $n=1.58$ for case D results in $\gamma=3.42$ which satisfies the constraint of NANOGrav collaboration for spectral index  as $\gamma=3.2\pm0.6$ \citep{NANOGrav-1,NANOGrav-2,NANOGrav-3,NANOGrav-4,NANOGrav-5}.
\end{itemize}

\section{Acknowledgements}\label{sec7}
The authors thank the referee for his/her valuable comments.


\end{document}